\documentclass[a4paper,12pt]{article}
\usepackage[utf8x]{inputenc}
\usepackage{amsfonts}
\usepackage{latexsym}
\usepackage{amsmath}
\usepackage{amssymb}
\usepackage{amssymb}

\usepackage{slashed}
\usepackage{upgreek}
\usepackage{mathrsfs}
\usepackage{bbm}

\usepackage{relsize}
\usepackage{graphicx}

\numberwithin{equation}{section}

\newcommand{\newsection}{    
						\setcounter{equation}{0}
						\section
					}
\def\appendix#1{	\addtocounter{section}{1}
				\setcounter{equation}{0}
				\renewcommand{\thesection}{\Alph{section}}
				\section*{Appendix \thesection\protect\indent \parbox[t]{11.15cm}{#1}}
				\addcontentsline{toc}{section}{Appendix \thesection\ \ \ #1}
			}

\newcommand{\be}{\begin{eqnarray}}
\newcommand{\ee}{\end{eqnarray}}
\newcommand{\bea}{\begin{eqnarray}}
\newcommand{\eea}{\end{eqnarray}}
\newcommand{\ba}{\begin{array}}
\newcommand{\ea}{\end{array}}

\newenvironment{frcseries}{\fontfamily{frc}\selectfont}{}

\def \la {\label}

\def\e{\epsilon}

\def\bbe{{\bf{e}}}
\font\mybb=msbm10 at 11pt

\def\bb#1{\hbox{\mybb#1}}

\def\bN{\bb{N}}
\def\bR {\bb{R}}

\def\tn {{\tilde{\nabla}}}

\begin{document}
\begin{titlepage}
\begin{center}
\vspace*{-1.0cm}
\hfill DMUS--MP--14/13 \\

\vspace{2.0cm} {\Large \bf Dynamical symmetry enhancement near IIA horizons} \\[.2cm]

\vspace{1.5cm}
 {\large  U.~Gran$^1$, J.~Gutowski$^2$, U.~Kayani$^3$ and  G.Papadopoulos$^3$}

\vspace{0.5cm}

${}^1$ Fundamental Physics\\
Chalmers University of Technology\\
SE-412 96 G\"oteborg, Sweden\\

\vspace{0.5cm}
$^2$ Department of Mathematics \\
University of Surrey \\
Guildford, GU2 7XH, UK \\

\vspace{0.5cm}
${}^3$ Department of Mathematics\\
King's College London\\
Strand\\
London WC2R 2LS, UK\\

\vspace{0.5cm}

\end{center}

\vskip 1.5 cm
\begin{abstract}

We show that smooth type IIA Killing horizons with compact spatial sections  preserve an even number of supersymmetries, and that the symmetry algebra of horizons with non-trivial fluxes includes an  $\mathfrak{sl}(2, \bR)$ subalgebra.
This confirms the conjecture of \cite{iibindex} for type IIA horizons. As an intermediate step in the proof, we also demonstrate  new Lichnerowicz type theorems
for spin bundle connections whose holonomy is contained in a general linear group.

\end{abstract}

\end{titlepage}



\setcounter{section}{0}
\setcounter{subsection}{0}
\setcounter{equation}{0}

\newsection{Introduction}

It has been conjectured in \cite{iibindex}, following earlier work in \cite{5index} and \cite{11index},  that
\begin{itemize}

\item the number of Killing spinors $N$, $N\not=0$,  of Killing horizons in supergravity is given by
\bea
N = 2N_{-} + \mathrm{Index}(D_{E})
\la{index}
\eea
where $N_{-}\in \bN_{>0}$ and $D_{E}$ is a Dirac operator twisted by a vector bundle $E$, defined on the spatial horizon section $\mathcal{S}$, which depends on the gauge symmetries of the supergravity theory in question,  and

\item that horizons with non-trivial fluxes and $N_{-} \neq 0$ admit an $\mathfrak{sl}(2,\bR)$ symmetry subalgebra.
\end{itemize}

This conjecture encompasses the essential features of (super)symmetry enhancement near black hole Killing horizons, and some features of the same phenomenon near brane horizons, previously obtained in the literature based on a case-by-case investigation
 \cite{carter, gibbons1, gwgpkt}. Symmetry enhancement near black hole and brane horizons has been instrumental in the development
 of the AdS/CFT correspondence \cite{maldacena}.
So far, this conjecture has been established for minimal 5-dimensional gauged supergravity,  D=11 M-theory,  and D=10 IIB supergravity \cite{5index, iibindex, 11index}.

The main purpose of this paper is to prove the above conjecture for Killing horizons in IIA supergravity. The proof is based on three assumptions. First, it is assumed  that the Killing horizons
admit at least one supersymmetry, second  that the near  horizon geometries are smooth and third that the spatial horizon  sections are compact without boundary\footnote{This is not
an essential assumption and it may be weakened. However to extend our proof to horizons with non-compact ${\cal S}$, one has to impose appropriate boundary
 conditions on the fields. Because of this, and for simplicity, we shall not do this here and throughout this paper we shall assume that ${\cal S}$ is compact without boundary.}.
It turns out that for IIA horizons, the contribution from the index of $D_E$ in the expression for $N$ in (\ref{index}) vanishes and therefore one concludes that IIA horizons always preserve an even number of supersymmetries, i.e.
\bea
N=2N_-~.
\eea
Furthermore from the second part of the conjecture, one concludes that all supersymmetric IIA horizons with non-trivial fluxes admit an $\mathfrak{sl}(2,\bR)$ symmetry subalgebra.

To prove the conjecture, we first adapt the description of black hole near horizon geometries of \cite{isen, gnull} to IIA supergravity. The metric and the remaining  fields
of IIA horizons are given in (\ref{hormetr}).  We then decompose the Killing spinor as $\e=\e_++\e_-$ using the lightcone projectors $\Gamma_\pm \e_\pm=0$ and integrate the Killing spinor equations (KSEs) of IIA supergravity along the two  lightcone directions. These directions  arise naturally in the description
of near horizon geometries. As a result, the Killing spinors of IIA horizons can be written as $\epsilon=\epsilon(u,r, \eta_\pm)$, where the dependence on the
coordinates $u,r$ is explicit and $\eta_\pm$ are spinors which depend only on the coordinates of the spatial horizon section ${\cal S}$ given by the equation $u=r=0$.

As a key next step in the proof,  we demonstrate that the remaining independent KSEs are those obtained from the KSEs of IIA supergravity after naively restricting them to ${\cal S}$.
In particular, we find after an extensive use of the field equations and Bianchi identities that all the integrability conditions that arise
along the lightcone directions,  and the mixed directions between the lightcone and the ${\cal S}$ directions, are automatically satisfied.
The independent KSEs on ${\cal S}$ split into two sets $\{\nabla^{(\pm)}, {\cal A}^{(\pm)}\}$ of two KSEs with each set acting on the spinors $\eta_\pm$ distinguished by the choice
of lightcone direction, where $\nabla^{(\pm)}$ are derived
 from the gravitino KSE of IIA supergravity and ${\cal A}^{(\pm)}$ are associated to the dilatino KSE of IIA supergravity. In addition we demonstrate that
if $\eta_-$ is a Killing spinor on ${\cal S}$, then $\eta_+=\Gamma_+ \Theta_- \eta_-$ also solves the KSEs, where $\Theta_-$ depends on the fluxes and the spacetime metric.

To show that the number of Killing spinors of IIA horizons is even, it suffices to show that there are as many $\eta_+$ Killing spinors as $\eta_-$ Killing spinors.
For this, we first identify the Killing spinors $\eta_\pm$ with the zero modes of  Dirac-like operators ${\mathscr D}^{(\pm)}$ coupled to fluxes. These are defined
as  ${\mathscr D}^{(\pm)}={\cal D}^{(\pm)}+ q{\cal A}^{(\pm)}$, where ${\cal D}^{(\pm)}$ is the Dirac operator constructed from $\nabla^{(\pm)}$.
It is then shown that for a suitable choice of $q$ all zero modes of these Dirac-like operators are in 1-1 correspondence with the Killing spinors.

The proof of the above correspondence between zero modes and Killing spinors for the ${\mathscr D}^{(+)}$ operator utilizes the Hopf maximum principle
and relies on the formula (\ref{maxprin}). Incidentally, this
also establishes that $\parallel \eta_+\parallel$ is constant.
The proof for the ${\mathscr D}^{(-)}$ operator uses the partial integration of the formula (\ref{l2b}) and this is similar to the classical
Lichnerowicz theorem for the Dirac operator.
 In both cases, the proofs rely on the smoothness of data and the assumption that
${\cal S}$ is compact without boundary.

Therefore, the number of Killing spinors of IIA horizons is $N=N_++N_-$,  where $N_\pm$ are the dimensions of the kernels of the ${\mathscr D}^{(\pm)}$ operators.
On the other hand, one can show that the zero modes of ${\mathscr D}^{(-)}$ are in 1-1 correspondence with the zero modes of the adjoint  $({\mathscr D}^{(+)})^\dagger$ of ${\mathscr D}^{(+)}$.
As a result $N_+-N_-$ is the index of ${\mathscr D}^{(+)}$. This vanishes as it is equal to the index of the Dirac operator
acting on the spinor bundle constructed from the  ${\bf 16}$ dimensional Majorana representation of $Spin(8)$. As a result $N_+=N_-$ and the number of supersymmetries
preserved by IIA horizons is even, which proves the first part of the conjecture.

To prove that IIA horizons admit an  $\mathfrak{sl}(2,\bR)$ symmetry subalgebra, we use the fact that if $\eta_-$ is a Killing spinor then $\eta_+=\Gamma_+\Theta_-\eta_-$
is also a Killing spinor. To see this we demonstrate that if the fluxes do not vanish, the kernel of $\Theta_-$ is $\{0\}$, and so $\eta_+\not=0$.
Using the Killing spinors now constructed from $\eta_-$ and $\eta_+=\Gamma_+\Theta_-\eta_-$, we prove that the spacetime admits three Killing vectors, which leave all the fields invariant, satisfying
 an $\mathfrak{sl}(2,\bR)$ algebra. This completes the proof of the conjecture for IIA horizons.

The results presented above for horizons in IIA supergravity do not follow from those we have obtained for M-horizons in \cite{11index}.  Although
IIA supergravity is the dimensional reduction of 11-dimensional supergravity, the reduction, after truncation of   Kaluza-Klein modes, does not
always preserve all the supersymmetry of 11-dimensional solutions; for a detailed analysis of these issues see \cite{bakas, duff}.
As a result, for example, it does not follow that IIA horizons preserve an even number of supersymmetries
because   M-horizons do as shown in \cite{11index}.  However since we have shown that both IIA and M-theory horizons preserve an even number of supersymmetries, one concludes
that if the reduction process breaks some supersymmetry, then it always breaks an even number of supersymmetries.

 This paper is organized as follows. In section 2, we identify the independent KSEs for IIA horizons. In section 3, we establish the equivalence between zero modes
 of  ${\mathscr D}^{(\pm)}$ and Killing spinors, and show that the number of supersymmetries preserved by IIA horizons is even. In section 4, we show that
 $\eta_+=\Gamma_+\Theta_-\eta_-\not=0$. In section 5, we prove that IIA horizons with non-trivial fluxes admit an $\mathfrak{sl}(2,\bR)$ symmetry subalgebra and in section 6 we give our conclusions. In appendix A, we give a list of Bianchi identities and field equations that are implied by the (independent) ones listed in section 2. In appendix B, we identify the independent KSEs, and in appendix C we
 establish the formulae (\ref{maxprin}) and (\ref{l2b}).

\newsection{Horizon fields and KSEs}

\subsection{IIA fields and field equations}

The bosonic field content  of IIA supergravity \cite{huq, giani, west, romans} are the spacetime metric  $g$, the dilaton $\Phi$, the 2-form NS-NS gauge potential $B$,
and the 1-form and the 3-form RR gauge potentials  $A$ and $C$, respectively. In addition,
the  theory has  non-chiral fermionic fields consisting of a Majorana gravitino and  a Majorana dilatino  but these are set to zero
in all the computations that follow.
The bosonic field strengths of IIA supergravity in the conventions of \cite{howe} are
\bea
F=dA~,~~~H=dB~,~~~G=dC-H\wedge A~.
\eea
These lead to the Bianchi identities
\bea
\label{bian1}
dF=0~,~~~dH=0~,~~~dG=F\wedge H~.
\eea
The bosonic part of the IIA action in the string frame is
\bea
S = \int \sqrt{-g} \bigg(e^{-2 \Phi} \big(R+4 \nabla_\mu \Phi\nabla^\mu \Phi -{1 \over 12} H_{\lambda_1 \lambda_2 \lambda_3}
H^{\lambda_1 \lambda_2 \lambda_3} \big)
\nonumber \\
-{1 \over 4} F_{\mu \nu} F^{\mu \nu}
-{1 \over 48} G_{\mu_1 \mu_2 \mu_3 \mu_4}
G^{\mu_1 \mu_2 \mu_3 \mu_4}\bigg) + {1 \over 2} dC \wedge dC \wedge B~.
\eea
This leads to the Einstein equation
\bea
\label{eineq}
R_{\mu \nu}&=&-2 \nabla_\mu \nabla_\nu \Phi
+{1 \over 4} H_{\mu \lambda_1 \lambda_2} H_\nu{}^{\lambda_1 \lambda_2}
+{1 \over 2} e^{2 \Phi} F_{\mu \lambda} F_\nu{}^\lambda
+{1 \over 12} e^{2 \Phi} G_{\mu \lambda_1 \lambda_2 \lambda_3}
G_\nu{}^{\lambda_1 \lambda_2 \lambda_3}
\nonumber \\
&+& g_{\mu \nu} \bigg(-{1 \over 8}e^{2 \Phi}
F_{\lambda_1 \lambda_2}F^{\lambda_1 \lambda_2}
-{1 \over 96}e^{2 \Phi} G_{\lambda_1 \lambda_2 \lambda_3
\lambda_4} G^{\lambda_1 \lambda_2 \lambda_3
\lambda_4} \bigg)~,
\eea
the dilaton field equation
\bea
\label{dileq}
\nabla^\mu \nabla_\mu \Phi
&=& 2 \nabla_\lambda \Phi \nabla^\lambda \Phi
-{1 \over 12} H_{\lambda_1 \lambda_2 \lambda_3}
H^{\lambda_1 \lambda_2 \lambda_3}+{3 \over 8} e^{2 \Phi}
F_{\lambda_1 \lambda_2} F^{\lambda_1 \lambda_2}
\nonumber \\
&+&{1 \over 96} e^{2 \Phi} G_{\lambda_1 \lambda_2 \lambda_3
\lambda_4} G^{\lambda_1 \lambda_2 \lambda_3
\lambda_4}~,
\eea
the 2-form field equation
\bea
\label{geq1}
\nabla^\mu F_{\mu \nu} +{1 \over 6} H^{\lambda_1 \lambda_2 \lambda_3} G_{\lambda_1 \lambda_2 \lambda_3 \nu} =0~,
\eea
the 3-form field equation
\bea
\label{beq1}
\nabla_\lambda \bigg( e^{-2 \Phi} H^{\lambda \mu \nu}\bigg) -{1 \over 2} G^{\mu \nu \lambda_1 \lambda_2} F_{\lambda_1 \lambda_2}
+{1 \over 1152} \epsilon^{\mu \nu \lambda_1 \lambda_2
\lambda_3 \lambda_4 \lambda_5 \lambda_6 \lambda_7 \lambda_8}
G_{\lambda_1 \lambda_2 \lambda_3 \lambda_4}
G_{\lambda_5 \lambda_6 \lambda_7 \lambda_8} =0~,
\eea
and the 4-form field equation
\bea
\label{ceq1}
\nabla_\mu G^{\mu \nu_1 \nu_2 \nu_3}
+{1 \over 144} \epsilon^{\nu_1 \nu_2 \nu_3
\lambda_1 \lambda_2 \lambda_3 \lambda_4 \lambda_5
\lambda_6 \lambda_7} G_{\lambda_1 \lambda_2 \lambda_3
\lambda_4} H_{\lambda_5 \lambda_6 \lambda_7}=0~.
\eea
This completes the description of the dynamics of the bosonic part of IIA supergravity.

\subsection{Horizon fields, Bianchi identities and field equations}

The description of the metric near extreme Killing horizons as expressed in Gaussian null coordinates \cite{isen, gnull}
can be adapted to include all IIA fields. In particular, one writes
\be
ds^2 &=&2 \bbe^+ \bbe^- + \delta_{ij} \bbe^i \bbe^j~,~~~G= \bbe^+ \wedge \bbe^- \wedge X +r \bbe^+ \wedge Y + \tilde G~,~~
\cr
H &=&\bbe^+ \wedge \bbe^- \wedge L+ r \bbe^+ \wedge M + \tilde H~,~~~F= \bbe^+ \wedge \bbe^- S + r \bbe^+ \wedge T+ \tilde F~,
\la{hormetr}
\ee
where we have introduced the frame
\be
\label{basis1}
\bbe^+ = du, \qquad \bbe^- = dr + rh -{1 \over 2} r^2 \Delta du, \qquad \bbe^i = e^i_I dy^I~,
\ee
and the dependence on the coordinates $u$ and $r$ is explicitly given. Moreover $\Phi$ and $\Delta$ are 0-forms, $h$, $L$ and $T$ are 1-forms, $X$, $M$ and $\tilde F$ are 2-forms,
$Y, \tilde H$ are 3-forms and $\tilde G$ is a 4-form on  the spatial horizon section ${\cal S}$, which is the co-dimension 2 submanifold given by the equation  $r=u=0$, i.e.~all these
components of the fields depend only on the coordinates of ${\cal S}$. It should be noted that one of our assumptions is that
all these forms on ${\cal S}$ are sufficiently differentiable, i.e.~we require at least $C^2$ differentiability
so that all the field equations and Bianchi identities are valid.

Substituting the fields (\ref{hormetr}) into the Bianchi identities  of IIA supergravity, one finds that
\bea
\label{bian2}
M&=&d_h L~,~~~T=d_h S~,~~~Y=d_h X-L\wedge \tilde F-S \tilde H~,~~~
\cr
d\tilde G&=&\tilde H\wedge \tilde F~,~~~d\tilde H=d\tilde F=0~,
\eea
where  $d_h \theta \equiv d \theta-
h \wedge \theta$ for any form $\theta$. These are the only independent Bianchi identities, see appendix A.

Similarly, substituting the horizon fields into the field equations of IIA supergravity,  we find that
the 2-form field equation ({\ref{geq1}) gives
\bea
\label{feq1}
\tn^i {\tilde{F}}_{ik}-h^i {\tilde{F}}_{ik}
+T_k -L^i X_{ik} +{1 \over 6} {\tilde{H}}^{\ell_1 \ell_2
\ell_3} {\tilde{G}}_{\ell_1 \ell_2 \ell_3 k}=0~,
\eea
the 3-form field equation  ({\ref{beq1}}) gives
\bea
\label{feq2}
\tn^i (e^{-2 \Phi}L_i) -{1 \over 2} {\tilde{F}}^{ij}
X_{ij} +{1 \over 1152}
\epsilon^{\ell_1 \ell_2 \ell_3 \ell_4 \ell_5 \ell_6 \ell_7
\ell_8} {\tilde{G}}_{\ell_1 \ell_2 \ell_3 \ell_4}
{\tilde{G}}_{\ell_5 \ell_6 \ell_7 \ell_8}=0
\eea
and
\bea
\label{feq3}
\tn^i(e^{-2 \Phi} {\tilde{H}}_{imn})
-e^{-2 \Phi} h^i {\tilde{H}}_{imn}
+e^{-2 \Phi} M_{mn}+S X_{mn}-{1 \over 2}
{\tilde{F}}^{ij} {\tilde{G}}_{ijmn}
\nonumber \\
-{1 \over 48} \epsilon_{mn}{}^{\ell_1
\ell_2 \ell_3 \ell_4 \ell_5 \ell_6} X_{\ell_1 \ell_2}
{\tilde{G}}_{\ell_3 \ell_4 \ell_5 \ell_6} =0~,
\eea
and the 4-form field equation ({\ref{ceq1}}) gives
\bea
\label{feq4}
\tn^i X_{ik} +{1 \over 144} \epsilon_k{}^{\ell_1 \ell_2
\ell_3 \ell_4 \ell_5 \ell_6 \ell_7}
{\tilde{G}}_{\ell_1 \ell_2 \ell_3 \ell_4} {\tilde{H}}_{\ell_5 \ell_6 \ell_7} =0
\eea
and
\bea
\label{feq5}
\tn^i {\tilde{G}}_{ijkq}+Y_{jkq}-h^i {\tilde{G}}_{ijkq}
-{1 \over 12} \epsilon_{jkq}{}^{\ell_1 \ell_2 \ell_3
\ell_4 \ell_5} X_{\ell_1 \ell_2} {\tilde{H}}_{\ell_3 \ell_4 \ell_5}
-{1 \over 24}\epsilon_{jkq}{}^{\ell_1 \ell_2 \ell_3
\ell_4 \ell_5} {\tilde{G}}_{\ell_1 \ell_2 \ell_3 \ell_4}
L_{\ell_5} =0~,
\nonumber \\
\eea
where $\tilde \nabla$ is the Levi-Civita connection of the metric on ${\cal S}$.
In addition, the dilaton field equation ({\ref{dileq}}) becomes
\bea
\label{feq6}
\tn^i \tn_i \Phi - h^i \tn_i \Phi &=&
2 \tn_i \Phi \tn^i \Phi +{1 \over 2} L_i L^i
-{1 \over 12} {\tilde{H}}_{\ell_1 \ell_2 \ell_3}
{\tilde{H}}^{\ell_1 \ell_2 \ell_3}-{3 \over 4} e^{2 \Phi}S^2
\nonumber \\
&+&{3 \over 8} e^{2 \Phi} {\tilde{F}}_{ij}
{\tilde{F}}^{ij} -{1 \over 8} e^{2 \Phi} X_{ij}X^{ij}
+{1 \over 96} e^{2 \Phi} {\tilde{G}}_{\ell_1 \ell_2 \ell_3
\ell_4} {\tilde{G}}^{\ell_1 \ell_2 \ell_3 \ell_4}~.
\eea
It remains to evaluate the Einstein field equation. This gives
\bea
\label{feq7}
{1 \over 2} \tn^i h_i -\Delta -{1 \over 2}h^2
&=& h^i \tn_i \Phi -{1 \over 2} L_i L^i -{1 \over 4} e^{2 \Phi} S^2 -{1 \over 8} e^{2 \Phi} X_{ij} X^{ij}
\nonumber \\
&-&{1 \over 8} e^{2 \Phi} {\tilde{F}}_{ij} {\tilde{F}}^{ij}
-{1 \over 96} e^{2 \Phi} {\tilde{G}}_{\ell_1 \ell_2
\ell_3 \ell_4} {\tilde{G}}^{\ell_1 \ell_2 \ell_3 \ell_4}~,
\eea
and
\bea
\label{feq8}
{\tilde{R}}_{ij} &=& -\tn_{(i} h_{j)}
+{1 \over 2} h_i h_j -2 \tn_i \tn_j \Phi
-{1 \over 2} L_i L_j +{1 \over 4} {\tilde{H}}_{i
\ell_1 \ell_2} {\tilde{H}}_j{}^{\ell_1 \ell_2}
\nonumber \\
&+&{1 \over 2} e^{2 \Phi} {\tilde{F}}_{i \ell}
{\tilde{F}}_j{}^\ell -{1 \over 2} e^{2 \Phi} X_{i \ell}
X_j{}^\ell+{1 \over 12} e^{2 \Phi} {\tilde{G}}_{i \ell_1 \ell_2 \ell_3} {\tilde{G}}_j{}^{\ell_1 \ell_2 \ell_3}
\nonumber \\
&+&\delta_{ij} \bigg({1 \over 4} e^{2 \Phi} S^2
-{1 \over 8} e^{2 \Phi} {\tilde{F}}_{\ell_1 \ell_2}
{\tilde{F}}^{\ell_1 \ell_2} +{1 \over 8} e^{2 \Phi}
X_{\ell_1 \ell_2}X^{\ell_1 \ell_2}
-{1 \over 96} e^{2 \Phi} {\tilde{G}}_{\ell_1 \ell_2
\ell_3 \ell_4} {\tilde{G}}^{\ell_1 \ell_2 \ell_3 \ell_4}
\bigg)~.
\nonumber \\
\eea
Above we have only stated the independent field equations. In fact, after substituting the near horizon geometries into the IIA field equations, there are additional  equations that arise.
However, these are all implied from the above field equations and Bianchi identities. For completeness, these additional equations are given in appendix A.

To summarize, the independent Bianchi identities and field equations are given in
 ({\ref{bian2}})--({\ref{feq8}}).

\subsection{Integration of KSEs along the lightcone}\label{ikse}

The KSEs of IIA supergravity are the vanishing conditions of the
gravitino and dilatino supersymmetry variations evaluated at the locus where
all fermions vanish.
These can be expressed as
\bea
{\cal D}_\mu\e&\equiv&\nabla_\mu\e +{1\over8} H_{\mu\nu_1\nu_2} \Gamma^{\nu_1\nu_2}\Gamma_{11}\e+{1\over16} e^\Phi F_{\nu_1\nu_2} \Gamma^{\nu_1\nu_2} \Gamma_\mu \Gamma_{11} \e
\cr
&& +{1\over 8\cdot 4!}e^\Phi G_{\nu_1\nu_2\nu_3\nu_4} \Gamma^{\nu_1\nu_2\nu_3\nu_4} \Gamma_\mu\e=0~,\label{GKSE}\\
{\cal A}\e&\equiv&\partial_\mu\Phi\, \Gamma^\mu \e+{1\over12} H_{\mu_1\mu_2\mu_3} \Gamma^{\mu_1\mu_2\mu_3} \Gamma_{11} \e+{3\over8} e^\Phi F_{\mu_1\mu_2} \Gamma^{\mu_1\mu_2} \Gamma_{11} \e
\cr
&&+ {1\over 4\cdot 4!}e^\Phi\, G_{\mu_1\mu_2\mu_3\mu_4} \Gamma^{\mu_1\mu_2\mu_3\mu_4}\e=0~,\label{AKSE}
\eea
where $\epsilon$ is the supersymmetry parameter which from now on is taken to be a Majorana, but not Weyl, commuting spinor of $Spin(9,1)$. In what follows, we shall refer
to the ${\cal D}$ operator as the supercovariant connection.

Supersymmetric IIA horizons are those for which there exists an $\epsilon\not=0$ that is a solution of the KSEs. To find the conditions
on the fields required for such a solution to exist, we first integrate along the two lightcone directions, i.e.~we integrate the KSEs
along the $u$ and $r$ coordinates. To do this, we decompose $\epsilon$ as
\bea
\e=\e_++\e_-~,
\label{ksp1}
\eea
 where $\Gamma_\pm\epsilon_\pm=0$, and find that
\bea\label{lightconesol}
\e_+=\phi_+(u,y)~,~~~\e_-=\phi_-+r \Gamma_-\Theta_+ \phi_+~,
\eea
and
\bea
\phi_-=\eta_-~,~~~\phi_+=\eta_++ u \Gamma_+ \Theta_-\eta_-~,
\eea
where
\bea
\Theta_\pm={1\over4} h_i\Gamma^i\mp{1\over4} \Gamma_{11} L_i \Gamma^i-{1\over16} e^{\Phi} \Gamma_{11} (\pm 2 S+\tilde F_{ij} \Gamma^{ij})-{1\over8 \cdot 4!} e^{\Phi} (\pm 12 X_{ij} \Gamma^{ij}
+\tilde G_{ijkl} \Gamma^{ijkl})~,
\nonumber \\
\eea
and $\eta_\pm$ depend only on the coordinates of the spatial horizon section ${\cal S}$. As spinors on ${\cal S}$, $\eta_\pm$ are sections of the $Spin(8)$ bundle on ${\cal S}$
associated with the Majorana representation.  Equivalently, the $Spin(9,1)$ bundle $S$ on the spacetime when restricted to ${\cal S}$ decomposes
as $S=S_-\oplus S_+$ according to the lightcone projections $\Gamma_\pm$. Although $S_\pm$ are distinguished by the lightcone chirality, they are isomorphic
as $Spin(8)$ bundles over ${\cal S}$. We shall use this in the counting of supersymmetries of IIA horizons.

\subsection{Independent KSEs}

The substitution of the  spinor (\ref{ksp1}) into the KSEs produces a large number of additional conditions.  These can be seen
either as integrability conditions along the lightcone directions, as well as integrability conditions along the mixed lightcone and
${\cal S}$ directions, or as KSEs along ${\cal S}$. A detailed analysis, presented in appendix \ref{indkse}, of the formulae obtained reveals
that the independent KSEs are those that are obtained from the naive restriction of the IIA KSEs to ${\cal S}$.  In particular,
the independent KSEs are
\bea
\label{covr}
\nabla_{i}^{(\pm)}\eta_{\pm}  = 0~,~~~\mathcal{A}^{(\pm)}\eta_{\pm} = 0~,
\eea
where
\bea
\nabla_{i}^{(\pm)}&=& \tilde{\nabla}_{i} + \Psi^{(\pm)}_{i}~,
\eea
with
\bea
\label{alg1pm}
\Psi^{(\pm)}_{i} &=& \bigg( \mp \frac{1}{4}h_{i} \mp \frac{1}{16}e^{\Phi}X_{l_1 l_2}\Gamma^{l_1 l_2}\Gamma_{i} + \frac{1}{8.4!}e^{\Phi}{\tilde{G}}_{l_1 l_2 l_3 l_4}\Gamma^{l_1 l_2 l_3 l_4}\Gamma_{i}\bigg)
\cr
&+& \Gamma_{11}\bigg(\mp \frac{1}{4}L_{i} + \frac{1}{8}{\tilde{H}}_{i l_1 l_2}\Gamma^{l_1 l_2}
\pm \frac{1}{8}e^{\Phi}S\Gamma_{i} - \frac{1}{16}e^{\Phi}{\tilde{F}}_{l_1 l_2}\Gamma^{l_1 l_2}\Gamma_{i}\bigg)~,
\eea
and
\bea
\label{alg2pm}
\mathcal{A}^{(\pm)} &=& \partial_i \Phi \Gamma^i  + \bigg(\mp \frac{1}{8}e^{\Phi}X_{l_1 l_2}\Gamma^{l_1 l_2} + \frac{1}{4.4!}e^{\Phi}{\tilde{G}}_{l_1 l_2 l_3 l_4}\Gamma^{l_1 l_2 l_3 l_4}\bigg)
\cr
&+& \Gamma_{11}\bigg(\pm \frac{1}{2}L_{i}\Gamma^{i} - \frac{1}{12}{\tilde{H}}_{i j k}\Gamma^{i j k} \mp \frac{3}{4}e^{\Phi}S + \frac{3}{8}e^{\Phi}{\tilde{F}}_{i j}\Gamma^{i j}\bigg)~.
\eea
Evidently, $\nabla^{(\pm)}$ arise from the supercovariant connection while $\mathcal{A}^{(\pm)}$ arise
from the dilatino KSE of IIA supergravity as restricted to ${\cal S}$ .

Furthermore, the analysis in appendix \ref{indkse} reveals that  if $\eta_{-}$ solves $(\ref{covr})$ then
\bea
\eta_+ = \Gamma_{+}\Theta_{-}\eta_{-}~,
\label{epfem}
\eea
 also solves $(\ref{covr})$. This is the first indication that IIA horizons admit an even number of supersymmetries.  As we shall prove, the existence of the $\eta_+$ solution
 is also responsible for the $\mathfrak{sl}(2,\bR)$ symmetry of IIA horizons.

\newsection{Supersymmetry enhancement}

To prove that IIA horizons always admit an even number of supersymmetries, it suffices to prove that there are as many $\eta_+$ Killing spinors as there are $\eta_-$ Killing spinors,
i.e.~that the $\eta_+$ and $\eta_-$ Killing spinors come in pairs. For this, we shall identify the Killing spinors with the zero modes of  Dirac-like operators
which depend on the fluxes and then use the index theorem to count their modes.

\subsection{Horizon Dirac equations}

We define  horizon Dirac operators associated with the supercovariant derivatives following from the gravitino KSE as
\bea
{\cal D}^{(\pm)} \equiv \Gamma^{i}\nabla_{i}^{(\pm)} = \Gamma^{i}\tilde{\nabla}_{i} + \Psi^{(\pm)}~,
\eea
where
\bea
\label{alg3pm}
\Psi^{(\pm)} \equiv \Gamma^{i}\Psi^{(\pm)}_{i} &=& \mp\frac{1}{4}h_{i}\Gamma^{i}
\mp\frac{1}{4}e^{\Phi}X_{i j}\Gamma^{i j}
\cr
&+& \Gamma_{11}\bigg(\pm \frac{1}{4}L_{i}\Gamma^{i} - \frac{1}{8}{\tilde{H}}_{i j k}\Gamma^{i j k} \mp e^{\Phi}S + \frac{1}{4}e^{\Phi}{\tilde{F}}_{i j}\Gamma^{i j} \bigg)~.
\eea
However, it turns out that
it is not possible to straightforwardly formulate Lichnerowicz theorems
to identify zero modes of these horizon Dirac operators with Killing spinors.

To proceed, we shall modify both the KSEs and the horizon Dirac operators.  For this first observe that an equivalent set of KSEs can be chosen
by redefining the supercovariant derivatives from the gravitino KSE as
\bea
\label{redef1}
\hat{\nabla}_{i}^{(\pm)}=\nabla_{i}^{(\pm)}+ \kappa \Gamma_i {\cal A}^{(\pm)}~,
\eea
for some $\kappa\in \bR$, because
\bea
\hat{\nabla}_{i}^{(\pm)}\eta_\pm=0~,~~~{\cal A}^{(\pm)}\eta_\pm=0\Longleftrightarrow {\nabla}_{i}^{(\pm)}\eta_\pm=0~,~~~{\cal A}^{(\pm)}\eta_\pm=0~.
\eea

Similarly, one can modify the horizon Dirac operators as
\bea
\label{redef2}
{\mathscr D}^{(\pm)}={\cal D}^{(\pm)} + q{\cal A}^{(\pm)}~,
\eea
for some $q\in \bR$. Clearly, if $q=8\kappa$, then ${\mathscr D}^{(\pm)}= \Gamma^i \hat{\nabla}_{i}^{(\pm)}$. However, we
shall not assume this in general. As we shall see, there is an appropriate choice of $q$ and appropriate choices of $\kappa$ such that
the Killing spinors can be identified with the zero modes of ${\mathscr D}^{(\pm)}$.

\subsection{A Lichnerowicz type theorem for $\mathcal{D}^{(+)}$}

First let us establish that the $\eta_+$ Killing spinors can be identified with the zero modes of a ${\mathscr D}^{(+)}$.  It is straightforward to see
that if $\eta_+$ is a Killing spinor, then $\eta_+$ is a zero mode of ${\mathscr D}^{(+)}$. So it remains to demonstrate the converse.
For this assume that $\eta_+$ is a zero mode of ${\mathscr D}^{(+)}$, i.e.~${\mathscr D}^{(+)}\eta_+=0$. Then after some lengthy computation which utilizes the field equations and Bianchi identities, described
in appendix C, one can establish the equality
\bea
{\tilde{\nabla}}^{i}{\tilde{\nabla}}_{i}\parallel\eta_+\parallel^2 - (2\tilde{\nabla}^i \Phi +  h^i) {\tilde{\nabla}}_{i}\parallel\eta_+\parallel^2 = 2\parallel{\hat\nabla^{(+)}}\eta_{+}\parallel^2 + (-4\kappa - 16 \kappa^2)\parallel\mathcal{A}^{(+)}\eta_+\parallel^2~,
\label{maxprin}
\eea
provided that $q=-1$. It is clear that if the last term on the right-hand-side of the above identity is positive semi-definite, then one can apply the maximum principle on $\parallel\eta_+\parallel^2$
as the fields are assumed to be smooth, and ${\cal S}$ compact.
In particular,  if
\bea
-{1\over4} <\kappa<0~,
\label{rangek}
\eea
then the maximum principle implies that $\eta_+$ are Killing spinors and $\parallel\eta_+\parallel=\mathrm{const}$. Observe that if one takes ${\mathscr D}^{(+)}$ with $q=-1$, then ${\mathscr D}^{(+)}= \Gamma^i \hat{\nabla}_{i}^{(+)}$ provided that
$\kappa=-1/8$ which lies in the range (\ref{rangek}).

To summarize we have established that for $q=-1$ and $-{1 \over 4} < \kappa <0$,
\bea
\nabla_{i}^{(+)}\eta_+=0~,~~~{\cal A}^{(+)}\eta_+=0~\Longleftrightarrow~ {\mathscr D}^{(+)}\eta_+=0~.
\eea
Moreover $\parallel\eta_+\parallel^2$ is constant on ${\cal S}$.

\subsection{A Lichnerowicz type theorem for $\mathcal{D}^{(-)}$}

Next we shall establish that the $\eta_-$ Killing spinors can also be identified with the zero modes of a modified horizon Dirac operator ${\mathscr D}^{(-)}$.
It is clear that all Killing spinors $\eta_-$ are zero modes of ${\mathscr D}^{(-)}$. To prove the converse,
suppose that $\eta_-$ satisfies ${\mathscr D}^{(-)} \eta_-=0$.
The proof proceeds by calculating the Laplacian of $\parallel \eta_-
\parallel^2$ as described in appendix C, which requires the use of the
field equations and Bianchi identies.
One can then establish the formula
\bea
\label{l2b}
{\tilde{\nabla}}^{i} \big( e^{-2 \Phi} V_i \big)
= -2 e^{-2 \Phi} \parallel{\hat\nabla^{(-)}}\eta_{-}\parallel^2 +   e^{-2 \Phi} (4 \kappa +16 \kappa^2) \parallel\mathcal{A}^{(-)}\eta_-\parallel^2~,
\nonumber \\
\eea
provided that $q=-1$, where
\bea
V=-d \parallel \eta_- \parallel^2 - \parallel \eta_- \parallel^2 h \ .
\eea
The last term on the RHS of ({\ref{l2b}}) is negative semi-definite
if $
-{1 \over 4} < \kappa <0$.
Provided that this holds, on integrating ({\ref{l2b}}) over
${\cal{S}}$ and assuming that ${\cal{S}}$ is compact and without boundary, one finds that ${\hat\nabla^{(-)}}\eta_{-}=0$ and $\mathcal{A}^{(-)}\eta_-=0$.

Therefore, we have shown that for $q=-1$ and $-{1 \over 4} < \kappa <0$,
\bea
\nabla_{i}^{(-)}\eta_-=0~,~~~{\cal A}^{(-)}\eta_-=0~\Longleftrightarrow~ {\mathscr D}^{(-)}\eta_-=0~ \ .
\eea
This concludes the relationship between Killing spinors and zero modes of modified horizon Dirac operators.

\subsection{Supersymmetry enhancement}

The analysis developed so far suffices to prove that IIA horizons preserve an even number of supersymmetries. Indeed, if $N_\pm$ is the number
of $\eta_\pm$ Killing spinors, then the number of supersymmetries of IIA horizon is $N=N_++N_-$. Utilizing the relation between the Killing spinors $\eta_\pm$
and the zero modes of the modified horizon Dirac operators ${\mathscr D}^{(\pm)}$ established in the previous two sections, we have that
\bea
N_\pm=\mathrm{dim}\,\mathrm{Ker}\, {\mathscr D}^{(\pm)}~.
\eea

Next let us focus on the index of the ${\mathscr D}^{(+)}$ operator. As we have mentioned, the spin bundle of the spacetime $S$ decomposes
on ${\cal S}$ as $S=S_+\oplus S_-$. Moreover, $S_+$ and $S_-$ are isomorphic as $Spin(8)$ bundles and are associated with the Majorana
non-Weyl ${\bf 16}$ representation. Furthermore ${\mathscr D}^{(+)}: \Gamma(S_+)\rightarrow \Gamma(S_+)$, where $ \Gamma(S_+)$ are the sections of $S_+$
and this action does not preserve the $Spin(8)$ chirality.  Since the principal symbol of ${\mathscr D}^{(+)}$ is the same as the principal symbol
of the standard Dirac operator acting on Majorana but not-Weyl spinors, the index vanishes\footnote{This should be contrasted to IIB horizons  where the horizon Dirac operators
 act on the Weyl spinors and map them to anti-Weyl ones.  As a result, the horizon Dirac operators have the same principal symbol as the standard Dirac operator acting on the Weyl spinors
 and so there is a non-trivial contribution from the index.} \cite{index}.  As a result, we conclude that
\bea
\mathrm{dim}\,\mathrm{Ker}\, {\mathscr D}^{(+)}= \mathrm{dim}\,\mathrm{Ker}\, ({\mathscr D}^{(+)})^\dagger~,
\eea
where $({\mathscr D}^{(+)})^\dagger$ is the adjoint of ${\mathscr D}^{(+)}$.  Furthermore observe that
\bea
\big(e^{2 \Phi} \Gamma_-\big) \big({\mathscr D}^{(+)}\big)^\dagger
= {\mathscr D}^{(-)} \big(e^{2 \Phi} \Gamma_-\big), \qquad  ({\rm for} \ q=-1)~,
\eea
and so
\bea
N_-=\mathrm{dim}\,\mathrm{Ker}\, ({\mathscr D}^{(-)})=\mathrm{dim}\,\mathrm{Ker}\, ({\mathscr D}^{(+)})^\dagger~.
\eea
Therefore, we conclude that $N_+=N_-$ and so the number of supersymmetries of IIA horizons $N=N_++N_-=2 N_-$ is even. This proves
the first part of the conjecture ({\ref{index}}) for IIA horizons.

\section{Construction of $\eta_+$ from $\eta_{-}$ Killing spinors}

In the investigation of the integrability conditions of the KSEs, we have demonstrated that if $\eta_-$ is a Killing spinor, then
$\eta_+ = \Gamma_+ \Theta_- \eta_-$ is also a Killing spinor, see (\ref{epfem}). Since we know that the $\eta_+$ and $\eta_-$ Killing spinors
appear in pairs, the formula (\ref{epfem}) provides a way to construct the $\eta_+$ Killing spinors from the $\eta_-$ ones.
However, this is the case provided that $\eta_+=\Gamma_+ \Theta_- \eta_-\not=0$. Here, we shall prove that for horizons with non-trivial fluxes
\bea
\mathrm{Ker}\, \Theta_-=\{0\}~,
\label{kerz}
\eea
 and so
the operator $\Gamma_+\Theta_-$ pairs the $\eta_-$ with the $\eta_+$ Killing spinors.

We shall prove  $\mathrm{Ker}\, \Theta_-=\{0\}$ using contradiction.  For this assume that $\Theta_-$ has a non-trivial kernel, i.e.~there is $\eta_-\not=0$
such that
\bea
\Theta_- \eta_-=0~.
\eea
If this is the case, then the last integrability condition in  (\ref{int1}) gives that
\begin{eqnarray}
\langle \eta_- ,  \bigg( -{1 \over 2} \Delta -{1 \over 8} dh_{ij} \Gamma^{ij} + \frac{1}{8}M_{i j}\Gamma^{i j}\Gamma_{11} - \frac{1}{4}e^{\Phi}T_{i}\Gamma^{i}\Gamma_{11} - \frac{1}{24}e^{\Phi}Y_{i j k}\Gamma^{i j k}\bigg)  \eta_- \rangle =0~.
\end{eqnarray}
This in turn implies that
\begin{eqnarray}
\Delta \langle \eta_- , \eta_- \rangle =0~,
\end{eqnarray}
and hence
\begin{eqnarray}
\Delta =0 \ ,
\end{eqnarray}
as $\eta_-$ is no-where vanishing.

Next the gravitino KSE $\nabla^{(-)}\eta_-=0$ implies that
\begin{eqnarray}
{\tilde{\nabla}}_i \langle \eta_-, \eta_- \rangle &=& -{1 \over 2} h_i  \langle \eta_-, \eta_- \rangle
+ \langle \eta_- , \bigg(\frac{1}{4}e^{\Phi}X_{i \ell}\Gamma^{\ell} - \frac{1}{96}e^{\Phi}\tilde{G}_{\ell_1 \ell_2 \ell_3 \ell_4}\Gamma_i{}^{\ell_1 \ell_2 \ell_3 \ell_4}\bigg) \eta_- \rangle~
\nonumber \\
&+& \langle \eta_{-}, \Gamma_{11}\bigg(-\frac{1}{2}L_{i}
+ \frac{1}{8}e^{\Phi}\tilde{F}_{\ell_1 \ell_2}\Gamma_i{}^{\ell_1 \ell_2} \bigg) \eta_{-} \rangle~,
\end{eqnarray}
which can be simplified further using
\bea
\langle \eta_{-}, \Gamma_{i}\Theta_{-}\eta_{-} \rangle &=& \frac{1}{4}h_{i} \langle \eta_-, \eta_- \rangle
+ \langle \eta_- , \bigg(\frac{1}{8}e^{\Phi}X_{i \ell}\Gamma^{\ell} - \frac{1}{192}e^{\Phi}\tilde{G}_{\ell_1 \ell_2 \ell_3 \ell_4}\Gamma_i{}^{\ell_1 \ell_2 \ell_3 \ell_4}\bigg) \eta_- \rangle~
\nonumber \\
&+& \langle \eta_{-}, \Gamma_{11}\bigg(-\frac{1}{4}L_{i}
+ \frac{1}{16}e^{\Phi}\tilde{F}_{\ell_1 \ell_2}\Gamma_i{}^{\ell_1 \ell_2} \bigg) \eta_{-} \rangle = 0~,
\eea
to yield
\begin{eqnarray}
\label{nrm1a}
{\tilde{\nabla}}_i \parallel \eta_-\parallel^2 = - h_i  \parallel \eta_-\parallel^2~.
\end{eqnarray}
As $\eta_-$ is no-where zero, this implies that
\begin{eqnarray}
dh=0~.
\end{eqnarray}
Substituting,  $\Delta=0$  and $dh=0$ into (\ref{auxeq4}), we  find that
\begin{eqnarray}
M=d_h L=0~,~~~T=d_h S=0~,~~~Y=d_h X-L\wedge \tilde F-S \tilde H=0~,
\end{eqnarray}
as well.
Returning to ({\ref{nrm1a}}), on taking the divergence, and using ({\ref{feq7}}) to
eliminate the ${\tilde{\nabla}}^i h_i$ term, one obtains
\begin{eqnarray}
{\tilde{\nabla}}^i {\tilde{\nabla}}_i  \parallel \eta_-\parallel^2 = 2\tilde{\nabla}^{i}\Phi {\tilde{\nabla}}_i \parallel \eta_-\parallel^2+ \bigg(L^2+ \frac{1}{2}e^{2\Phi}S^2 + \frac{1}{4}e^{2\Phi}X^2 + \frac{1}{4}e^{2\Phi}\tilde{F}^2 + \frac{1}{48}e^{2\Phi}\tilde{G}^2\bigg)  \parallel \eta_-\parallel^2~.
\nonumber \\
\end{eqnarray}
Applying the maximum principle on $\parallel \eta_-\parallel^2$ we conclude that all the fluxes apart from the dilaton $\Phi$ and $\tilde H$ vanish and
$\parallel\eta_-\parallel$ is constant. The latter together with  (\ref{nrm1a}) imply that $h=0$.

Next applying the maximum principle to the dilaton field equation (\ref{feq6}), we conclude that the dilaton is constant and $\tilde H=0$.
Combining all the results so far, we conclude that all the fluxes vanish which is a contradiction to the assumption that not all of the fluxes vanish.  This establishes (\ref{kerz}).

Furthermore, the horizons for which $\Theta_- \eta_- =0$ ($\eta_- \neq 0$) are all local products  $\bR^{1,1}\times {\cal S}$,  where ${\cal S}$ up to a discrete identification is a product of Ricci flat Berger manifolds. Thus ${\cal S}$ has holonomy, $Spin(7)$ or $SU(4)$ or $Sp(2)$ as an  irreducible manifold,  and $G_2$ or $SU(3)$ or $Sp(1)\times Sp(1)$ or $Sp(1)$ or $\{1\}$ as a reducible one.

\newsection{The $\mathfrak{sl}(2,\bR)$ symmetry of IIA horizons}

It remains to prove the second part of the conjecture that all IIA horizons with non-trivial fluxes admit an $\mathfrak{sl}(2,\bR)$ symmetry subalgebra.
As we shall demonstrate, this in fact is a consequence of our previous result that all IIA horizons admit an even number of supersymmetries. The  proof is
 very similar to that already given in the context of M-horizons in \cite{11index}, so we shall be brief.

\subsection{Killing vectors}

To begin, first note that the Killing spinor $\epsilon$ on the spacetime can be expressed in terms of $\eta_\pm$ as
\bea
\epsilon= \eta_++ u \Gamma_+\Theta_-\eta_-+ \eta_-+r \Gamma_-\Theta_+\eta_++ru \Gamma_-\Theta_+\Gamma_+\Theta_-\eta_-~,
\label{gensolkse}
\eea
which is derived after collecting the results of section \ref{ikse}.

Since the $\eta_-$ and $\eta_+$ Killing spinors appear in pairs for supersymmetric IIA horizons, let us choose a $\eta_-$ Killing spinor.  Then from the results
of the previous section, horizons with non-trivial fluxes also admit $\eta_+=\Gamma_+\Theta_-\eta_-$ as a Killing spinors. Using $\eta_-$ and $\eta_+=\Gamma_+\Theta_-\eta_-$,
one can construct two linearly independent Killing spinors on the  spacetime as
\bea
\epsilon_1=\eta_-+u\eta_++ru \Gamma_-\Theta_+\eta_+~,~~~\epsilon_2=\eta_++r\Gamma_-\Theta_+\eta_+~.
\eea

To continue, it is known from the general theory of supersymmetric IIA backgrounds that for any Killing spinors $\zeta_1$ and $\zeta_2$ the dual vector field of the 1-form
bilinear
\bea
K(\zeta_1, \zeta_2)=\langle(\Gamma_+-\Gamma_-) \zeta_1, \Gamma_a\zeta_2\rangle\, e^a~,
\eea
is a Killing vector and leaves invariant all the other fields of the theory.
Evaluating, the 1-form bilinears of the Killing spinor $\epsilon_1$ and $\epsilon_2$, we find that
\begin{eqnarray}
 K_1(\epsilon_1, \epsilon_2)&=& (2r \langle\Gamma_+\eta_-, \Theta_+\eta_+\rangle+  u^2 r \Delta \parallel \eta_+\parallel^2) \,{\bf{e}}^+-2u \parallel\eta_+\parallel^2\, {\bf{e}}^-+ V_i {\bf{e}}^i~,
 \cr
 K_2(\epsilon_2, \epsilon_2)&=& r^2 \Delta\parallel\eta_+\parallel^2 \,{\bf{e}}^+-2 \parallel\eta_+\parallel^2 {\bf{e}}^-~,
 \cr
 K_3(\epsilon_1, \epsilon_1)&=&(2\parallel\eta_-\parallel^2+4r u \langle\Gamma_+\eta_-, \Theta_+\eta_+\rangle+ r^2 u^2 \Delta \parallel\eta_+\parallel^2) {\bf{e}}^+
 \cr
 && \qquad\qquad \qquad\qquad -2u^2 \parallel\eta_+\parallel^2 {\bf{e}}^-+2u V_i {\bf{e}}^i~,
 \label{b1forms}
 \end{eqnarray}
where we have set
\begin{eqnarray}
\label{vii}
V_i =  \langle \Gamma_+ \eta_- , \Gamma_i \eta_+ \rangle~.
\end{eqnarray}
Moreover, we have used the identities
\begin{eqnarray}
- \Delta\, \parallel\eta_+\parallel^2 +4  \parallel\Theta_+ \eta_+\parallel^2 =0~,~~~\langle \eta_+ , \Gamma_i \Theta_+ \eta_+ \rangle  =0~,
\end{eqnarray}
which follow from the first integrability condition in ({\ref{int1}}),  $\parallel\eta_+\parallel=\mathrm{const}$ and the KSEs of $\eta_+$.

\subsection{The geometry of ${\cal S}$}

First suppose that $V\not=0$. Then the conditions ${\cal L}_{K_a} g=0$ and ${\cal L}_{K_a} F=0$, $a=1,2,3$, where $F$ denotes collectively all the
fluxes of IIA supergravity, imply that
\begin{eqnarray}
\tilde\nabla_{(i} V_{j)}=0~,~~~\tilde {\cal L}_Vh=\tilde {\cal L}_V\Delta=0~,~~~ \tilde {\cal L}_V \Phi=0~,
\nonumber \\
\tilde {\cal L}_V X=\tilde {\cal L}_V \tilde G=\tilde {\cal L}_V L=\tilde {\cal L}_V \tilde H=
\tilde {\cal L}_V S=\tilde {\cal L}_V \tilde F=0~,
\end{eqnarray}
i.e.~$V$ is an isometry of ${\cal S}$ and leaves all the fluxes on ${\cal S}$ invariant.
In addition, one also finds the useful identities
\begin{eqnarray}
&&-2 \parallel\phi_+\parallel^2-h_i V^i+2 \langle\Gamma_+\phi_-, \Theta_+\phi_+\rangle=0~,~~~i_V (dh)+2 d \langle\Gamma_+\phi_-, \Theta_+\phi_+\rangle=0~,
\cr
&& 2 \langle\Gamma_+\phi_-, \Theta_+\phi_+\rangle-\Delta \parallel\phi_-\parallel^2=0~,~~~
V+ \parallel\phi_-\parallel^2 h+d \parallel\phi_-\parallel^2=0~,
\label{conconx}
\end{eqnarray}
which imply that ${\cal L}_V\parallel\phi_-\parallel^2=0$. There are further restrictions on the geometry of ${\cal S}$ which will be explored elsewhere.

A special case arises for $V=0$ where the group action generated by $K_1, K_2$ and $K_3$ has only 2-dimensional orbits. A direct substitution of this condition in (\ref{conconx}) reveals that
\begin{eqnarray}
\Delta \parallel\phi_-\parallel^2=2 \parallel\phi_+\parallel^2~,~~~h=\Delta^{-1} d\Delta~.
\end{eqnarray}
Since $dh=0$ and $h$ is exact such horizons are static and a coordinate transformation $r\rightarrow \Delta r$ reveals that the horizon geometry is a warped product of $AdS_2$ with ${\cal S}$, $AdS_2\times_w {\cal S}$.

\subsection{$\mathfrak{sl}(2,\mathbb{R})$ symmetry of IIA-horizons}

To uncover the $\mathfrak{sl}(2,\mathbb{R})$ symmetry of IIA horizons it remains to compute the Lie bracket algebra of the vector fields associated to the 1-forms $K_1, K_2$ and $K_3$.
For this note that these vector fields can be expressed as
\begin{eqnarray}
K_1&=&-2u \parallel\eta_+\parallel^2 \partial_u+ 2r \parallel\eta_+\parallel^2 \partial_r+ V^i \tilde \partial_i~,
\cr
K_2&=&-2 \parallel\eta_+\parallel^2 \partial_u~,
\cr
K_3&=&-2u^2 \parallel\eta_+\parallel^2 \partial_u +(2 \parallel\eta_-\parallel^2+ 4ru \parallel\eta_+\parallel^2)\partial_r+ 2u V^i \tilde \partial_i~,
\end{eqnarray}
where we have used the same symbol for the 1-forms and the associated vector fields. These expressions are
similar to those we have obtained for M-horizons in \cite{11index} apart form the range of the index $i$ which is different. Using the various identities we have obtained, a direct computation reveals
 that the Lie bracket algebra is
\begin{eqnarray}
[K_1,K_2]=2 \parallel\eta_+\parallel^2 K_2~,~~~[K_2, K_3]=-4 \parallel\eta_+\parallel^2 K_1  ~,~~~[K_3,K_1]=2 \parallel\eta_+\parallel^2 K_3~, \ \
\end{eqnarray}
which is isomorphic to $\mathfrak{sl}(2,\mathbb{R})$.
This proves the second part of the conjecture and completes the analysis.

\newsection{Conclusions}

We have demonstrated that smooth IIA horizons with compact spatial sections, without boundary,  always admit an even number of supersymmetries. In addition, those
with non-trivial fluxes admit an $\mathfrak{sl}(2,\mathbb{R})$ symmetry subalgebra.

The above result together with those obtained in \cite{5index, 11index} and \cite{iibindex} provide further evidence in support the conjecture
of \cite{iibindex} regarding the (super)symmetries of supergravity horizons. It also emphases that the (super)symmetry enhancement
that is observed near the horizons of supersymmetric black holes is a consequence of the smoothness of the fields.

Apart from exhibiting an $\mathfrak{sl}(2,\mathbb{R})$ symmetry, IIA horizons are further geometrically restricted.  This is because we have not explored
all the restrictions imposed by the KSEs and the field equations of the theory -- in this paper we only explored enough to establish the $\mathfrak{sl}(2,\mathbb{R})$ symmetry.
However, the understanding of the horizons admitting two supersymmetries is within the capability of the technology developed so far for the classification of supersymmetric IIA backgrounds \cite{iiacla} and it will be explored
elsewhere. The understanding of all IIA horizons is a more involved problem. As such spaces preserve an even number of supersymmetries and there are no IIA horizons
with non-trivial fluxes preserving 32 supersymmetries, which follows from the classification of maximally supersymmetric backgrounds in \cite{maxsusy}, there are potentially 15 different
cases to examine. Of course, all IIA horizons preserving more than 16 supersymmetries are homogenous spaces as a consequence of the results of \cite{josefof}. It is also very likely that there are no
IIA horizons preserving 28 and 30 supersymmetries in analogy with a similar result in IIB \cite{nearhor}.  However to prove this, it is required to extend the IIB classification
results   to IIA supergravity, see also \cite{josesp}.

We expect that our results on IIA horizons can be extended to massive IIA supergravity \cite{romans}. This will be reported elsewhere.

\vskip 1cm

\noindent{\bf Acknowledgements} \vskip 0.1cm
UG is supported by the Knut and Alice Wallenberg Foundation. GP is partially supported by the STFC grant ST/J002798/1. JG is supported by the STFC grant, ST/1004874/1.
JG would like to thank the
Department of Mathematical Sciences, University of Liverpool for hospitality during which part of this work
was completed. UK is supported by a STFC PhD fellowship.

\vskip 0.5cm


\setcounter{section}{0}
\setcounter{equation}{0}

\appendix{Horizon  Bianchi identities and field equations}
\label{bfe1}

We remark that there are a number of additional Bianchi identities, which are
\bea
dT + S dh + dS \wedge h &=&0~,
\nonumber \\
dM+L \wedge dh -h \wedge dL &=&0~,
\nonumber \\
dY + dh \wedge X - h \wedge dX + h \wedge (S {\tilde{H}}
+{\tilde{F}} \wedge L)+T \wedge {\tilde{H}}+{\tilde{F}} \wedge M &=&0~.
\eea
However, these Bianchi identities are implied by those in ({\ref{bian2}}).

There is also a number of additional field equations given by
\bea
\label{auxeq1}
-\tn^i T_i + h^i T_i -{1 \over 2} dh^{ij} {\tilde{F}}_{ij}
-{1 \over 2} X_{ij}M^{ij} -{1 \over 6} Y_{ijk}{\tilde{H}}^{ijk}=0~,
\eea
\bea
\label{auxeq2}
- \tn^i (e^{-2 \Phi}M_{ik}) + e^{-2 \Phi} h^i M_{ik}
-{1 \over 2} e^{-2 \Phi} dh^{ij} {\tilde{H}}_{ijk}
-T^i X_{ik} -{1 \over 2} {\tilde{F}}^{ij} Y_{ijk}
\nonumber \\
-{1 \over 144} \epsilon_k{}^{\ell_1
\ell_2 \ell_3 \ell_4 \ell_5 \ell_6 \ell_7}
Y_{\ell_1 \ell_2 \ell_3} {\tilde{G}}_{\ell_4 \ell_5 \ell_6 \ell_7} =0~,
\eea
\bea
\label{auxeq3}
- \tn^i Y_{imn}+h^i Y_{imn}-{1 \over 2} dh^{ij}
{\tilde{G}}_{ijmn}
+{1 \over 36} \epsilon_{mn}{}^{\ell_1 \ell_2 \ell_3
\ell_4 \ell_5 \ell_6} Y_{\ell_1 \ell_2 \ell_3}
{\tilde{H}}_{\ell_4 \ell_5 \ell_6}
\nonumber \\
+{1 \over 48} \epsilon_{mn}{}^{\ell_1 \ell_2 \ell_3
\ell_4 \ell_5 \ell_6} {\tilde{G}}_{\ell_1 \ell_2 \ell_3
\ell_4} M_{\ell_5 \ell_6}=0~,
\eea
corresponding to equations obtained from
the $+$ component of ({\ref{geq1}}),
the $k$ component of ({\ref{beq1}}) and
the $mn$ component of ({\ref{ceq1}}) respectively.
However, ({\ref{auxeq1}}), ({\ref{auxeq2}}) and
({\ref{auxeq3}}) are implied by ({\ref{feq1}})-
({\ref{feq5}}) together with the Bianchi identities
({\ref{bian2}}).

Note also that the $++$ and $+i$ components of the
Einstein equation, which are
\bea
\label{auxeq4}
{1 \over 2} \tn^i \tn_i \Delta -{3 \over 2} h^i \tn_i \Delta-{1 \over 2} \Delta \tn^i h_i
+ \Delta h^2 +{1 \over 4} dh_{ij} dh^{ij}
&=&(\tn^i \Delta - \Delta h^i)\tn_i \Phi +{1 \over 4} M_{ij}
M^{ij}
\nonumber \\
&+&{1 \over 2} e^{2 \Phi} T_i T^i
+{1 \over 12} e^{2 \Phi} Y_{ijk} Y^{ijk}
\nonumber \\
\eea
and
\bea
\label{auxeq5}
{1 \over 2} \tn^j dh_{ij}-dh_{ij} h^j - \tn_i \Delta + \Delta h_i
&=& dh_i{}^j \tn_j \Phi -{1 \over 2} M_i{}^j L_j
+{1 \over 4} M_{\ell_1 \ell_2} {\tilde{H}}_i{}^{\ell_1 \ell_2}-{1 \over 2} e^{2 \Phi} S T_i
\nonumber \\
&+&{1 \over 2} e^{2 \Phi} T^j {\tilde{F}}_{ij}
-{1 \over 4} e^{2 \Phi} Y_i{}^{\ell_1 \ell_2}
X_{\ell_1 \ell_2} +{1 \over 12} e^{2 \Phi}
Y_{\ell_1 \ell_2 \ell_3} {\tilde{G}}_i{}^{\ell_1 \ell_2
\ell_3}
\nonumber \\
\eea
are implied by ({\ref{feq6}}), ({\ref{feq7}}), ({\ref{feq8}}),
together with ({\ref{feq1}})-({\ref{feq5}}),
and the Bianchi identities ({\ref{bian2}}).

\newsection{Integrability conditions and KSEs} \label{indkse}

Substituting the solution of the KSEs along the lightcone directions (\ref{lightconesol})  back into the gravitino KSE (\ref{GKSE}) and appropriately expanding in the $r,u$ coordinates, we find that
for  the $\mu = \pm$ components, one obtains  the additional conditions
\bea
\label{int1}
&&\bigg({1\over2}\Delta - {1\over8}(dh)_{ij}\Gamma^{ij} + {1\over8}M_{ij}\Gamma_{11}\Gamma^{ij}
\cr
&&+ 2\big( {1\over 4} h_i \Gamma^{i} - {1\over 4} L_{i}\Gamma_{11}\Gamma^{i} - {1\over 16} e^{\Phi}\Gamma_{11}(-2S + \tilde{F}_{i j}\Gamma^{i j}) - \frac{1}{8\cdot4!}e^{\Phi}(12X_{i j}\Gamma^{i j} - \tilde{G}_{i j k l}\Gamma^{i j k l})\big )\Theta_{+} \bigg)\phi_{+} = 0
\nonumber \\
\eea
\bea
\label{int2}
&&\bigg(\frac{1}{4}\Delta h_i \Gamma^{i} - \frac{1}{4}\partial_{i}\Delta \Gamma^{i} + \big(-\frac{1}{8}(dh)_{ij}\Gamma^{ij} - \frac{1}{8}M_{ij}\Gamma^{ij}\Gamma_{11} - \frac{1}{4}e^{\Phi}T_{i}\Gamma^{i}\Gamma_{11} + \frac{1}{24}e^{\Phi}Y_{i j k}\Gamma^{i j k} \big) \Theta_{+} \bigg) \phi_{+} = 0
\nonumber \\
\eea
\bea
\label{int3}
&&\bigg(-\frac{1}{2}\Delta - \frac{1}{8}(dh)_{ij}\Gamma^{ij} + \frac{1}{8}M_{ij}\Gamma^{ij}\Gamma_{11} - \frac{1}{4}e^{\Phi} T_{i}\Gamma^{i} \Gamma_{11} - \frac{1}{24}e^{\Phi}Y_{ijk}\Gamma^{ijk}
\cr
&&+ 2\big( -{1\over4} h_i\Gamma^i-{1\over4} \Gamma_{11} L_i \Gamma^i +{1\over16} e^\phi \Gamma_{11} (2 S+\tilde F_{ij} \Gamma^{ij})-{1\over8 \cdot 4!} e^\phi (12 X_{ij} \Gamma^{ij}
+\tilde G_{ijkl} \Gamma^{ijkl}) \big) \Theta_{-} \bigg)\phi_{-} = 0 \ .
\nonumber \\
\eea

Similarly the $\mu=i$ component of the gravitino KSEs gives
\bea
\label{int4}
&&\tilde \nabla_i \phi_\pm\mp {1\over 4} h_i \phi_\pm \mp {1\over4} \Gamma_{11} L_i \phi_\pm +{1\over8} \Gamma_{11} \tilde H_{ijk} \Gamma^{jk} \phi_\pm
\cr
&&-{1\over16} e^\Phi \Gamma_{11} (\mp 2 S+ \tilde F_{kl} \Gamma^{kl}) \Gamma_i \phi_\pm+{1\over 8\cdot 4!} e^\Phi (\mp 12 X_{kl} \Gamma^{kl}+ \tilde G_{j_1j_2j_3j_4} \Gamma^{j_1j_2j_3j_4}) \Gamma_i \phi_\pm=0 \
\nonumber \\
\eea
and
\bea
\label{int5}
&&\tilde \nabla_i \tau_{+} + \bigg( -\frac{3}{4}h_i - \frac{1}{16}e^{\Phi}X_{l_1 l_2}\Gamma^{l_1 l_2}\Gamma_{i} - \frac{1}{8\cdot4!}e^{\Phi} \tilde G_{l_1\cdots l_4}\Gamma^{l_1 \cdots l_4}\Gamma_{i}
\cr
&&- \Gamma_{11}(\frac{1}{4}L_i + \frac{1}{8}\tilde{H}_{i j k}\Gamma^{j k} + \frac{1}{8}e^{\Phi} S \Gamma_{i} +    \frac{1}{16}e^{\Phi}\tilde{F}_{l_1 l_2}\Gamma^{l_1 l_2}\Gamma_{i})\bigg )\tau_{+}
\cr
&&+ \bigg(-\frac{1}{4}(dh)_{ij}\Gamma^{j} - \frac{1}{4}M_{ij}\Gamma^{j}\Gamma_{11} + \frac{1}{8}e^{\Phi}T_{j}\Gamma^{j}\Gamma_{i}\Gamma_{11} + \frac{1}{48}e^{\Phi}Y_{l_1 l_2 l_3}\Gamma^{l_1 l_2 l_3}\Gamma_{i} \bigg)\phi_{+} = 0
\nonumber \\
\eea
where we have set
\bea
\label{int6}
\tau_{+} = \Theta_{+}\phi_{+} \ .
\eea
All the additional conditions above can be viewed as integrability conditions along the lightcone and mixed lightcone and ${\cal S}$ directions.
We shall demonstrate that upon using the field equations and the Bianchi identities, the only independent conditions are (\ref{covr}).

\subsection{Dilatino KSE}

Substituting the solution of the KSEs (\ref{lightconesol})  into the dilatino KSE (\ref{AKSE}) and expanding appropriately in the $r,u$ coordinates, one obtains the following additional conditions

\bea
\label{int7}
&&\partial_i \Phi \Gamma^i \phi_{\pm} -{1\over12} \Gamma_{11} (\mp 6 L_i \Gamma^i+\tilde H_{ijk} \Gamma^{ijk}) \phi_\pm+{3\over8} e^\Phi \Gamma_{11} (\mp2 S+\tilde F_{ij} \Gamma^{ij})\phi_\pm
\cr
&&
+{1\over 4\cdot 4!}e^{\Phi} (\mp 12 X_{ij} \Gamma^{ij}+\tilde G_{j_1j_2j_3j_4} \Gamma^{j_1j_2j_3j_4}) \phi_\pm=0 \ ,
\eea
\be
\label{int8}
&&-\bigg( \partial_{i}\Phi\Gamma^{i} + \frac{1}{12}\Gamma_{11} (6L_i \Gamma^{i} + \tilde{H}_{ijk}\Gamma^{ijk}) + \frac{3}{8}e^{\Phi}\Gamma_{11}(2S + \tilde{F}_{ij}\Gamma^{ij})
\cr
&&- \frac{1}{4\cdot 4!}e^{\Phi}(12X_{ij}\Gamma^{ij} + \tilde{G}_{ijkl}\Gamma^{ijkl})\bigg)\tau_{+}
\cr
&&+ \bigg(\frac{1}{4}M_{ij}\Gamma^{ij}\Gamma_{11} + \frac{3}{4}e^{\Phi}T_{i}\Gamma^{i}\Gamma_{11} + \frac{1}{24}e^{\Phi}Y_{ijk}\Gamma^{ijk}\bigg)\phi_{+}=0~.
\eea
We shall show that the only independent ones are those in (\ref{covr}).

\subsection{Independent KSEs}
It is well known that the KSEs imply some of the Bianchi identities and field equations
of a theory. Because of this, to find solutions it is customary to solve the KSEs and then
impose the remaining field equations and Bianchi identities. However, we shall not do
this here because of the complexity of solving the KSEs (\ref{int1}), (\ref{int2}), (\ref{int5}), and (\ref{int8})
which contain the $\tau_+$ spinor as expressed in (\ref{int6}). Instead, we shall first show that all
the KSEs which contain $\tau_+$ are actually implied from those containing $\phi_+$, i.e. (\ref{int4}) and
(\ref{int7}), and some of the field equations and Bianchi identities.
Then we also show that (\ref{int3}) and the terms linear in u from the $+$ components of (\ref{int4}) and (\ref{int7}) are implied
by the field equations, Bianchi identities and the $-$ components of (\ref{int4}) and (\ref{int7}).

\subsubsection{The (\ref{int5}) condition}

The (\ref{int5}) component of the KSEs is implied by (\ref{int4}), (\ref{int6}) and (\ref{int7}) together with a number of field equations and Bianchi identities. First evaluate the LHS of (\ref{int5}) by substituting in (\ref{int6}) to eliminate $\tau_+$, and use (\ref{int4}) to evaluate the supercovariant derivative of $\phi_+$. Also, using (\ref{int4}) one can compute
\bea
&&(\tilde{\nabla}_{j}\tilde{\nabla}_{i} - \tilde{\nabla}_{i}\tilde{\nabla}_{j})\phi_{+}
 = {1\over 4} \tilde \nabla_j (h_i) \phi_+ + {1\over4} \Gamma_{11} \tilde \nabla_j(L_i) \phi_+ -{1\over8} \Gamma_{11} \tilde \nabla_j (\tilde H_{i l_1 l_2}) \Gamma^{l_1 l_2} \phi_+
\cr
&&+{1\over16} e^\Phi \Gamma_{11} (- 2\tilde \nabla_j (S)+ \tilde \nabla_j (\tilde F_{kl}) \Gamma^{kl}) \Gamma_i \phi_+ - {1\over 8\cdot 4!} e^\Phi (- 12 \tilde \nabla_j (X_{kl}) \Gamma^{kl}+ \tilde \nabla_j (\tilde G_{j_1j_2j_3j_4}) \Gamma^{j_1j_2j_3j_4}) \Gamma_i \phi_+
\cr
&&+{1\over16} \tilde \nabla_j \Phi e^\Phi \Gamma_{11} (- 2 S+  \tilde F_{kl} \Gamma^{kl}) \Gamma_i \phi_+ - {1\over 8\cdot 4!} \tilde \nabla_j \Phi e^\Phi (- 12  X_{kl} \Gamma^{kl}+  \tilde G_{j_1j_2j_3j_4} \Gamma^{j_1j_2j_3j_4}) \Gamma_i \phi_+
\cr
&&+ \big( {1\over 4} h_i  + {1\over4} \Gamma_{11} L_i  -{1\over8} \Gamma_{11} \tilde H_{ijk} \Gamma^{jk} +{1\over16} e^\Phi \Gamma_{11} (- 2 S+ \tilde F_{kl} \Gamma^{kl}) \Gamma_i
\cr
&&-{1\over 8\cdot 4!} e^\Phi (- 12 X_{kl} \Gamma^{kl}+ \tilde G_{j_1j_2j_3j_4} \Gamma^{j_1j_2j_3j_4}) \Gamma_i\big)\tilde \nabla_{j} \phi_+ - (i \leftrightarrow j) \ .
\label{DDphicond}
\eea
Then consider the following, where the first terms cancels from the definition of curvature,
\bea
\bigg(\frac{1}{4}\tilde{R}_{ij}\Gamma^{j} - \frac{1}{2}\Gamma^{j}(\tilde{\nabla}_{j}\tilde{\nabla}_{i} - \tilde{\nabla}_{i}\tilde{\nabla}_{j}) \bigg)\phi_+ + \frac{1}{2}\tilde{\nabla}_{i}(\mathcal{A}_1) + \frac{1}{2}\Psi_{i} \mathcal{A}_1 = 0~,
\label{B5intcond}
\eea
where
\bea
\mathcal{A}_1 &=& \partial_i \Phi \Gamma^i \phi_{+} -{1\over12} \Gamma_{11} (- 6 L_i \Gamma^i+\tilde H_{ijk} \Gamma^{ijk}) \phi_++{3\over8} e^\Phi \Gamma_{11} (-2 S+\tilde F_{ij} \Gamma^{ij})\phi_+
\cr
&
+&{1\over 4\cdot 4!}e^{\Phi} (- 12 X_{ij} \Gamma^{ij}+\tilde G_{j_1j_2j_3j_4} \Gamma^{j_1j_2j_3j_4}) \phi_+
\label{A1cond}
\eea
and
\bea
\Psi_{i} &=& - \frac{1}{4}h_{i} + \Gamma_{11}(\frac{1}{4}L_{i} - \frac{1}{8}\tilde{H}_{i j k}\Gamma^{j k})~.
\label{Psiicond}
\eea
The expression in (\ref{A1cond}) vanishes on making use of (\ref{int7}), as $\mathcal{A}_1 = 0$ is equivalent to the $+$ component of (\ref{int7}). However a non-trivial identity is obtained by using (\ref{DDphicond}) in (\ref{B5intcond}), and expanding out the $\mathcal{A}_1$ terms. Then, on adding (\ref{B5intcond}) to the LHS of (\ref{int5}), with $\tau_+$ eliminated in favour of $\eta_+$ as described above, one obtains the following
 \bea
&&\frac{1}{4}\bigg(\tilde{R}_{ij} + \tn_{(i} h_{j)}
-{1 \over 2} h_i h_j + 2 \tn_i \tn_j \Phi
+{1 \over 2} L_i L_j -{1 \over 4} {\tilde{H}}_{i
l_1 l_2} {\tilde{H}}_j{}^{l_1 l_2}
\cr
&-&{1 \over 2} e^{2 \Phi} {\tilde{F}}_{i l}
{\tilde{F}}_j{}^l + {1 \over 8} e^{2 \Phi} {\tilde{F}}_{l_1 l_2}
{\tilde{F}}^{l_1 l_2}\delta_{ij} + {1 \over 2} e^{2 \Phi} X_{i l}
X_j{}^l - {1 \over 8} e^{2 \Phi}
X_{l_1 l_2}X^{l_1 l_2}\delta_{ij}
\cr
&-&{1 \over 12} e^{2 \Phi} {\tilde{G}}_{i \ell_1 \ell_2 \ell_3} {\tilde{G}}_j{}^{\ell_1 \ell_2 \ell_3}
+ {1 \over 96} e^{2 \Phi} {\tilde{G}}_{\ell_1 \ell_2
\ell_3 \ell_4} {\tilde{G}}^{\ell_1 \ell_2 \ell_3 \ell_4}\delta_{ij} - {1 \over 4} e^{2 \Phi} S^2\delta_{ij}\bigg)\Gamma^{j}=0~.
\eea
This vanishes identically on making use of the Einstein equation (\ref{feq8}). Therefore it follows that (\ref{int5}) is implied by the $+$ component of (\ref{int4}), (\ref{int6}) and (\ref{int7}), the Bianchi identities (\ref{bian2}) and the gauge field equations (\ref{feq1})-(\ref{feq5}).

\subsubsection{The (\ref{int8}) condition}
Let us define
\bea
\mathcal{A}_2 =
&&-\bigg( \partial_{i}\Phi\Gamma^{i} + \frac{1}{12}\Gamma_{11} (6L_i \Gamma^{i} + \tilde{H}_{ijk}\Gamma^{ijk}) + \frac{3}{8}e^{\Phi}\Gamma_{11}(2S + \tilde{F}_{ij}\Gamma^{ij})
\cr
&&- \frac{1}{4\cdot 4!}e^{\Phi}(12X_{ij}\Gamma^{ij} + \tilde{G}_{ijkl}\Gamma^{ijkl})\bigg)\tau_{+}
\cr
&&+ \bigg(\frac{1}{4}M_{ij}\Gamma^{ij}\Gamma_{11} + \frac{3}{4}e^{\Phi}T_{i}\Gamma^{i}\Gamma_{11} + \frac{1}{24}e^{\Phi}Y_{ijk}\Gamma^{ijk}\bigg)\phi_{+}~,
\eea
where $\mathcal{A}_2$ equals the expression in (\ref{int8}).
One obtains the following identity
\bea
\mathcal{A}_2 = -\frac{1}{2}\Gamma^{i}\tilde{\nabla}_i \mathcal{A}_1 + \Psi_1\mathcal{A}_1 ~,
\eea
where
\bea
\Psi_1 &&= \tilde{\nabla}_{i}\Phi\Gamma^{i} + \frac{3}{8}h_{i}\Gamma^{i} + \frac{1}{16}e^{\Phi}X_{l_1 l_2}\Gamma^{l_1 l_2} - \frac{1}{192}e^{\Phi}\tilde{G}_{l_1 l_2 l_3 l_4}\Gamma^{l_1 l_2 l_3 l_4}
\cr
&&+ \Gamma_{11}\bigg(\frac{1}{48}\tilde{H}_{l_1 l_2 l_3}\Gamma^{l_1 l_2 l_3} - \frac{1}{8}L_{i}\Gamma^{i} + \frac{1}{16}e^{\Phi}\tilde{F}_{l_1 l_2}\Gamma^{l_1 l_2} - \frac{1}{8}e^{\Phi}S \bigg)~.
\eea
We have made use of the $+$ component of (\ref{int4}) in order to evaluate the covariant derivative in the above expression. In addition we have made use of the Bianchi identities (\ref{bian2}) and the field equations (\ref{feq1})-(\ref{feq6}).

\subsubsection{The (\ref{int1}) condition}
\label{B1sec}

In order to show that (\ref{int1}) is implied from the independent KSEs we can compute the following,
\bea
&&\bigg(-\frac{1}{4}\tilde{R} - \Gamma^{i j}\tilde{\nabla}_{i}\tilde{\nabla}_{j}\bigg)\phi_{+} - \Gamma^{i}\tilde{\nabla}_i(\mathcal{A}_1)
\cr
&+& \bigg(\tilde{\nabla}_i\Phi \Gamma^{i} + \frac{1}{4}h_{i}\Gamma^{i} + \frac{1}{16}e^{\Phi}X_{l_1 l_2}\Gamma^{l_1 l_2} - \frac{1}{192}e^{\Phi}\tilde{G}_{l_1 l_2 l_3 l_4}\Gamma^{l_1 l_2 l_3 l_4}
\cr
&+& \Gamma_{11}(-\frac{1}{4}L_{l}\Gamma^{l} - \frac{1}{24}\tilde{H}_{l_1 l_2 l_3}\Gamma^{l_1 l_2 l_3} - \frac{1}{8}e^{\Phi}S
+ \frac{1}{16}e^{\Phi}\tilde{F}_{l_1 l_2}\Gamma^{l_1 l_2}) \bigg)\mathcal{A}_1 = 0~,
\eea
where
\bea
\tilde{R} &&= -2\Delta - 2h^{i}\tilde{\nabla}_{i}\Phi - 2\tilde{\nabla}^2\Phi - \frac{1}{2}h^2 + \frac{1}{2}L^2 + \frac{1}{4}\tilde{H}^{2} + \frac{5}{2}e^{2\Phi}S^2
\cr
&&- \frac{1}{4}e^{2\Phi}\tilde{F}^2 + \frac{3}{4}e^{2\Phi}X^2 + \frac{1}{48}e^{2\Phi}\tilde{G}^2
\eea
and where we use the $+$ component of (\ref{int4}) to evaluate the covariant derivative terms. In order to obtain (\ref{int1}) from these expressions we make use of the Bianchi identities (\ref{bian2}), the field equations (\ref{feq1})-(\ref{feq6}), in particular in order to eliminate the $(\tilde{\nabla} \Phi)^2$ term. We have also made use of the $+-$ component of the Einstein equation (\ref{feq7}) in order to rewrite the scalar curvature $\tilde{R}$ in terms of $\Delta$. Therefore (\ref{int1}) follows from (\ref{int4}) and (\ref{int7}) together with the field equations and Bianchi identities mentioned above.

\subsubsection{The + (\ref{int7}) condition linear in $u$}
Since $\phi_+ = \eta_+ + u\Gamma_{+}\Theta_{-}\eta_-$, we must consider the part of the $+$ component of (\ref{int7}) which is linear in $u$. On defining
\bea
\mathcal{B}_1 &=& \partial_i \Phi \Gamma^i \eta_{-} -{1\over12} \Gamma_{11} ( 6 L_i \Gamma^i+\tilde H_{ijk} \Gamma^{ijk}) \eta_-+{3\over8} e^\Phi \Gamma_{11} (2 S+\tilde F_{ij} \Gamma^{ij})\eta_-
\cr
&&
+{1\over 4\cdot 4!}e^{\Phi} ( 12 X_{ij} \Gamma^{ij}+\tilde G_{j_1j_2j_3j_4} \Gamma^{j_1j_2j_3j_4}) \eta_-
\eea
one finds that the $u$-dependent part of (\ref{int7}) is proportional to
\bea
-\frac{1}{2}\Gamma^{i}\tilde{\nabla}_{i}(\mathcal{B}_1) + \Psi_2 \mathcal{B}_1~,
\eea
where
\bea
\Psi_2 &&= \tilde{\nabla}_{i}\Phi\Gamma^{i} + \frac{1}{8}h_{i}\Gamma^{i} - \frac{1}{16}e^{\Phi}X_{l_1 l_2}\Gamma^{l_1 l_2} - \frac{1}{192}e^{\Phi}\tilde{G}_{l_1 l_2 l_3 l_4}\Gamma^{l_1 l_2 l_3 l_4}
\cr
&&+ \Gamma_{11}\bigg(\frac{1}{48}\tilde{H}_{l_1 l_2 l_3}\Gamma^{l_1 l_2 l_3} + \frac{1}{8}L_{i}\Gamma^{i} + \frac{1}{16}e^{\Phi}\tilde{F}_{l_1 l_2}\Gamma^{l_1 l_2} + \frac{1}{8}e^{\Phi}S \bigg)~.
\eea
We have made use of the $-$ component of (\ref{int4}) in order to evaluate the covariant derivative in the above expression. In addition we have made use of the Bianchi identities (\ref{bian2}) and the field equations (\ref{feq1})-(\ref{feq6}).

\subsubsection{The (\ref{int2}) condition }
In order to show that (\ref{int2}) is implied from the independent KSEs we will show that it follows from (\ref{int1}). First act on (\ref{int1}) with the Dirac operator $\Gamma^{i}\tilde{\nabla}_{i}$ and use the field equations (\ref{feq1}) - (\ref{feq6}) and the Bianchi identities to eliminate the terms which contain derivatives of the fluxes and then use (\ref{int1}) to rewrite the $dh$-terms in terms of $\Delta$. Then use the conditions (\ref{int4}) and (\ref{int5}) to eliminate the $\partial_i \phi$-terms from the resulting expression, some of the remaining terms will vanish as a consequence of (\ref{int1}). After performing these calculations, the condition (\ref{int2}) is obtained, therefore it follows from section \ref{B1sec} above that (\ref{int2}) is implied by (\ref{int4}) and (\ref{int7}) together with the field equations and Bianchi identities mentioned above.

\subsubsection{The (\ref{int3}) condition }
In order to show that (\ref{int3}) is implied by the independent KSEs we can compute the following,
\bea
&&\bigg(\frac{1}{4}\tilde{R} + \Gamma^{i j}\tilde{\nabla}_{i}\tilde{\nabla}_{j}\bigg)\eta_- + \Gamma^{i}\tilde{\nabla}_i(\mathcal{B}_1)
\cr
&+& \bigg(-\tilde{\nabla}_i\Phi \Gamma^{i} + \frac{1}{4}h_{i}\Gamma^{i} + \frac{1}{16}e^{\Phi}X_{l_1 l_2}\Gamma^{l_1 l_2} + \frac{1}{192}e^{\Phi}\tilde{G}_{l_1 l_2 l_3 l_4}\Gamma^{l_1 l_2 l_3 l_4}
\cr
&+& \Gamma_{11}(-\frac{1}{4}L_{l}\Gamma^{l} + \frac{1}{24}\tilde{H}_{l_1 l_2 l_3}\Gamma^{l_1 l_2 l_3} - \frac{1}{8}e^{\Phi}S
- \frac{1}{16}e^{\Phi}\tilde{F}_{l_1 l_2}\Gamma^{l_1 l_2}) \bigg)\mathcal{B}_1 = 0~,
\eea
where we use the $-$ component of (\ref{int4}) to evaluate the covariant derivative terms. The expression above vanishes identically since the $-$ component of (\ref{int7}) is equivalent to $\mathcal{B}_1 = 0$. In order to obtain (\ref{int3}) from these expressions we make use of the Bianchi identities (\ref{bian2}) and the field equations (\ref{feq1})-(\ref{feq6}). Therefore (\ref{int3}) follows from (\ref{int4}) and (\ref{int7}) together with the field equations and Bianchi identities mentioned above.

\subsubsection{The + (\ref{int4}) condition linear in $u$}

Next consider the part of the $+$ component of (\ref{int4}) which is linear in $u$. First compute
\bea
\bigg(\Gamma^{j}(\tilde{\nabla}_{j}\tilde{\nabla}_{i} - \tilde{\nabla}_{i}\tilde{\nabla}_{j})  -\frac{1}{2}\tilde{R}_{ij}\Gamma^{j}\bigg)\eta_- - \tilde{\nabla}_{i}(\mathcal{B}_1) - \Psi_{i} \mathcal{B}_1 = 0~,
\eea
where
\bea
\Psi_{i} &=& \frac{1}{4}h_{i} - \Gamma_{11}(\frac{1}{4}L_{i} + \frac{1}{8}\tilde{H}_{i j k}\Gamma^{j k})
\eea
and where we have made use of the $-$ component of (\ref{int4}) to evaluate the covariant derivative terms. The resulting expression corresponds to the expression obtained by expanding out the $u$-dependent part of the $+$ component of (\ref{int4}) by using the $-$ component of (\ref{int4}) to evaluate the covariant derivative. We have made use of the Bianchi identities (\ref{bian2}) and the field equations (\ref{feq1})-(\ref{feq5}).


\appendix{Calculation of Laplacian of $\parallel \eta_\pm \parallel^2$}
\label{maxpex}

In this appendix, we calculate the Laplacian of $\parallel \eta_\pm \parallel^2$, which will be particularly useful in the analysis
of the global properties of IIA horizons in Section 3.
We shall consider the modified gravitino KSE ({\ref{redef1}})
defined in section 3.1, and we
shall assume throughout that the modified Dirac equation ${\mathscr D}^{(\pm)}\eta_\pm=0$ holds, where ${\mathscr D}^{(\pm)}$ is defined in
({\ref{redef2}}). Also, $\Psi^{(\pm)}_i$ and $\mathcal{A}^{(\pm)}$
are defined by ({\ref{alg1pm}}) and ({\ref{alg2pm}}), and
$\Psi^{(\pm)}$ is defined by ({\ref{alg3pm}}).

To proceed, we compute the Laplacian
\bea
\tilde{\nabla}^i \tilde{\nabla}_i ||\eta_{\pm}||^2 = 2\langle\eta_\pm,\tilde{\nabla}^i \tilde{\nabla}_i\eta_\pm\rangle + 2 \langle\tilde{\nabla}^i \eta_\pm, \tilde{\nabla}_i \eta_\pm\rangle \ .
\eea
To evaluate this expression note that
\bea
\tilde{\nabla}^i \tilde{\nabla}_i \eta_\pm &=& \Gamma^{i}\tilde{\nabla}_{i}(\Gamma^{j}\tilde{\nabla}_j \eta_\pm) -\Gamma^{i j}\tilde{\nabla}_i \tilde{\nabla}_j \eta_\pm
\nonumber \\
&=& \Gamma^{i}\tilde{\nabla}_{i}(\Gamma^{j}\tilde{\nabla}_j \eta_\pm) + \frac{1}{4}\tilde{R}\eta_\pm
\nonumber \\
&=& \Gamma^{i}\tilde{\nabla}_{i}(-\Psi^{(\pm)}\eta_\pm -q\mathcal{A}^{(\pm)}\eta_{\pm}) + \frac{1}{4}\tilde{R} \eta_\pm \ .
\eea
It follows that
\bea
\langle\eta_\pm,\tilde{\nabla}^i \tilde{\nabla}_i\eta_\pm \rangle &=& \frac{1}{4}\tilde{R}\parallel \eta_\pm \parallel^2
+ \langle\eta_\pm, \Gamma^{i}\tilde{\nabla}_i(-\Psi^{(\pm)} - q\mathcal{A}^{(\pm)})\eta_\pm\rangle
\nonumber \\
&+& \langle\eta_\pm, \Gamma^{i}(-\Psi^{(\pm)} - q\mathcal{A}^{(\pm)})\tilde{\nabla}_i \eta_\pm \rangle~,
\eea
and also
\bea
\langle\tilde{\nabla}^i \eta_\pm, \tilde{\nabla}_i \eta_\pm\rangle &=& \langle{\hat\nabla^{(\pm)i}} \eta_\pm, {\hat\nabla^{(\pm)}_{i}} \eta_\pm\rangle - 2\langle\eta_\pm, (\Psi^{(\pm)i} + \kappa\Gamma^{i}\mathcal{A}^{(\pm)})^{\dagger} \tilde{\nabla}_i \eta_\pm \rangle
\nonumber \\
&-& \langle\eta_\pm, (\Psi^{(\pm)i} + \kappa\Gamma^{i}\mathcal{A}^{(\pm)})^{\dagger} (\Psi^{(\pm)}_i + \kappa \Gamma_{i} \, \mathcal{A}^{(\pm)}) \eta_\pm \rangle
\nonumber \\
&=& \parallel {\hat\nabla^{(\pm)}}\eta_{\pm} \parallel^2 - 2\langle \eta_{\pm}, \Psi^{(\pm)i\dagger}\tilde{\nabla}_{i}\eta_{\pm}\rangle
- 2\kappa \langle \eta_{\pm}, \mathcal{A}^{(\pm)\dagger}\Gamma^{i}\tilde{\nabla}_{i}\eta_{\pm}\rangle
\nonumber \\
&-&  \langle \eta_\pm, (\Psi^{(\pm)i\dagger}\Psi^{(\pm)}_i + 2\kappa \mathcal{A}^{(\pm)\dagger}\Psi^{(\pm)} + 8\kappa^2\mathcal{A}^{(\pm)\dagger}\mathcal{A}^{(\pm)})\eta_\pm \rangle
\nonumber \\
&=& \parallel {\hat\nabla^{(\pm)}}\eta_{\pm} \parallel^2 - 2\langle \eta_{\pm}, \Psi^{(\pm)i\dagger}\tilde{\nabla}_{i}\eta_{\pm}\rangle - \langle \eta_\pm , \Psi^{(\pm)i\dagger}\Psi^{(\pm)}_i \eta_\pm \rangle
\nonumber \\
&+& (2\kappa q - 8\kappa^2)\parallel \mathcal{A}^{(\pm)}\eta_\pm \parallel^2 \ .
\eea
Therefore,
\bea
\label{extralap1b}
\frac{1}{2}\tilde{\nabla}^i \tilde{\nabla}_i ||\eta_{\pm}||^2 &=& \parallel {\hat\nabla^{(\pm)}}\eta_{\pm} \parallel^2 + \, (2\kappa q - 8\kappa^2)\parallel \mathcal{A}^{(\pm)}\eta_\pm \parallel^2
\nonumber \\
&+& \langle \eta_\pm, \bigg(\frac{1}{4}\tilde{R} + \Gamma^{i}\tilde{\nabla}_i(-\Psi^{(\pm)} - q\mathcal{A}^{(\pm)}) - \Psi^{(\pm)i\dagger}\Psi^{(\pm)}_i \bigg) \eta_\pm \rangle
\nonumber \\
&+& \langle \eta_\pm, \bigg( \Gamma^{i}(-\Psi^{(\pm)} - q\mathcal{A}^{(\pm)}) - 2\Psi^{(\pm)i\dagger}\bigg)\tilde{\nabla}_i \eta_\pm \rangle \ .
\eea
In order to simplify the expression for the Laplacian, we shall attempt to rewrite the third line in ({\ref{extralap1b}}) as
\bea
\label{bilin}
\langle \eta_\pm, \bigg( \Gamma^{i}(-\Psi^{(\pm)} - q\mathcal{A}^{(\pm)}) - 2\Psi^{(\pm)i\dagger}\bigg)\tilde{\nabla}_i \eta_\pm \rangle = \langle \eta_\pm, \mathcal{F}^{(\pm)}\Gamma^{i}\tilde{\nabla}_i \eta_\pm \rangle + W^{(\pm)i}\tilde{\nabla}_i \parallel \eta_\pm \parallel^2~,
\nonumber \\
\eea
where $\mathcal{F}^{(\pm)}$ is linear in the fields and $W^{(\pm)i}$ is a vector. This expression is particularly advantageous, because the
first term on the RHS can be rewritten using the horizon
Dirac equation, and the second term is consistent with the application
of the maximum principle/integration by parts arguments which
are required for the generalized Lichnerowicz theorems. In order to rewrite ({\ref{bilin}}) in this fashion, note that
\bea
\Gamma^{i}(\Psi^{(\pm)} + q\mathcal{A}^{(\pm)}) + 2\Psi^{(\pm)i\dagger} &=& \big(\mp h^i \mp (q+1)\Gamma_{11}L^{i} + {1 \over 2}(q+1)\Gamma_{11}\tilde{H}^{i}{}_{\ell_1 \ell_2}\Gamma^{\ell_1 \ell_2}
+ 2q\tilde{\nabla}^i \Phi \big)
\nonumber \\
&+& \big(\pm \frac{1}{4}h_{j}\Gamma^{j} \pm (\frac{q}{2} + \frac{1}{4})\Gamma_{11} L_{j}\Gamma^{j}
\nonumber \\
&-& (\frac{q}{12} + \frac{1}{8})\Gamma_{11}\tilde{H}_{\ell_1 \ell_2 \ell_3}\Gamma^{\ell_1 \ell_2 \ell_3}- q\tilde{\nabla}_j \Phi \Gamma^{j}\big)\Gamma^{i}
\nonumber \\
&\mp &{1 \over 8}(q+1)e^{\Phi}X_{\ell_1 \ell_2}\Gamma^{i}\Gamma^{\ell_1 \ell_2} +{1 \over 96} (q+1)e^{\Phi}\tilde{G}_{\ell_1 \ell_2 \ell_3 \ell_4}\Gamma^{i}\Gamma^{\ell_1 \ell_2 \ell_3 \ell_4}
\nonumber \\
&+& (q+1)\Gamma_{11}\bigg(\pm {3 \over 4} e^{\Phi}S\Gamma^{i}
-{3 \over 8}e^{\Phi}\tilde{F}_{\ell_1 \ell_2}\Gamma^{i}\Gamma^{\ell_1 \ell_2}\bigg) \ .
\eea
One finds that (\ref{bilin}) is only
possible for $q=-1$ and thus we have
\bea
W^{(\pm)i} = \frac{1}{2}(2\tilde{\nabla}^i \Phi \pm h^i)
\eea
\bea
\mathcal{F}^{(\pm)} = \mp \frac{1}{4}h_{j}\Gamma^{j} - \tilde{\nabla}_{j}\Phi \Gamma^{j} + \Gamma_{11}\bigg(\pm \frac{1}{4}L_{j}\Gamma^{j} +  \frac{1}{24}\tilde{H}_{\ell_1 \ell_2 \ell_3}\Gamma^{\ell_1 \ell_2 \ell_3}\bigg) \ .
\eea

We remark that  $\dagger$ is the adjoint with respect to the $Spin(8)$-invariant inner product $\langle \phantom{i},\phantom{i} \rangle$. In order to compute the adjoints above we note that the $Spin(8)$-invariant inner product restricted to the Majorana representation is positive definite and real, and so symmetric. With respect to this the gamma matrices are Hermitian and thus the skew symmetric products $\Gamma^{[k]}$ of $k$ $Spin(8)$ gamma matrices are Hermitian for $k=0 \, (\text{mod }4)$ and $k = 1 \, (\text{mod }4)$ while they are anti-Hermitian for $k = 2 \, (\text{mod }4)$ and $k = 3 \, (\text{mod }4)$. The $\Gamma_{11}$ matrix is also Hermitian since it is a product of the first 10 gamma matrices and we take $\Gamma_0$ to be anti-Hermitian. It also follows that $\Gamma_{11}\Gamma^{[k]}$ is Hermitian for $k=0\, (\text{mod }4)$ and $k=3\, (\text{mod }4)$ and anti-Hermitian for $k= 1 \, (\text{mod }4)$ and $k=2\, (\text{mod }4)$. This also implies the following identities
\bea
\label{hermiden1}
\langle \eta_+, \Gamma^{[k]} \eta_+ \rangle = 0, \qquad
k = 2\, (\text{mod }4) \ {\rm and} \  k=3\, (\text{mod }4)
\eea
and
\bea
\label{hermiden2}
\langle \eta_+, \Gamma_{11}\Gamma^{[k]} \eta_+ \rangle = 0,
\qquad k = 1\, (\text{mod }4) \ {\rm and} \  k= 2\, (\text{mod }4) \ .
\eea

It follows that
\bea
\label{laplacian}
\frac{1}{2}\tilde{\nabla}^i \tilde{\nabla}_i ||\eta_{\pm}||^2 &=& \parallel {\hat\nabla^{(\pm)}}\eta_{\pm} \parallel^2 + \, (-2\kappa  - 8\kappa^2)\parallel \mathcal{A}^{(\pm)}\eta_\pm \parallel^2
+ W^{(\pm)i}\tilde{\nabla}_{i}\parallel \eta_\pm \parallel^2
\nonumber \\
&+& \langle \eta_\pm, \bigg(\frac{1}{4}\tilde{R} + \Gamma^{i}\tilde{\nabla}_i(-\Psi^{(\pm)} + \mathcal{A}^{(\pm)}) - \Psi^{(\pm)i\dagger}\Psi^{(\pm)}_i  + \mathcal{F}^{(\pm)}(-\Psi^{(\pm)} + \mathcal{A}^{(\pm)})\bigg) \eta_\pm \rangle \ .
\nonumber \\
\eea
It is also useful to evaluate ${\tilde{R}}$ using (\ref{feq8}) and the dilaton field equation (\ref{feq6}); we obtain
\bea
\tilde{R} &=& -\tilde{\nabla}^{i}(h_i) + \frac{1}{2}h^2 - 4(\tilde{\nabla}\Phi)^2 - 2h^{i}\tilde{\nabla}_{i}\Phi - \frac{3}{2}L^2 + \frac{5}{12}\tilde{H}^2
\nonumber \\
&+& \frac{7}{2}e^{2\Phi}S^2 - \frac{5}{4}e^{2\Phi}\tilde{F}^2 + \frac{3}{4}e^{2\Phi}X^2 - \frac{1}{48}e^{2\Phi}\tilde{G}^2 \ .
\eea
One obtains, upon using the field equations and Bianchi identities,
\bea
\label{quad}
\bigg(\frac{1}{4}\tilde{R} &+& \Gamma^{i}\tilde{\nabla}_i(-\Psi^{(\pm)} + \mathcal{A}^{(\pm)}) - \Psi^{(\pm)i\dagger}\Psi^{(\pm)}_i  + \mathcal{F}^{(\pm)}(-\Psi^{(\pm)} + \mathcal{A}^{(\pm)})\bigg)\eta_\pm
\nonumber \\
&=& \bigg[ \big(\pm \frac{1}{4}\tilde{\nabla}_{\ell_1}(h_{\ell_2}) \mp \frac{1}{16}\tilde{H}^{i}{}_{\ell_1 \ell_2}L_{i}\big)\Gamma^{\ell_1 \ell_2}+  \big( \pm \frac{1}{8}\tilde{\nabla}_{\ell_1}(e^{\Phi}X_{\ell_2 \ell_3}) + \frac{1}{24}\tilde{\nabla}^{i}(e^{\Phi}\tilde{G}_{i \ell_1 \ell_2 \ell_3})
\nonumber \\
&\mp& \frac{1}{96}e^{\Phi}h^{i}\tilde{G}_{i \ell_1 \ell_2 \ell_3} -\frac{1}{32}e^{\Phi}X_{\ell_1 \ell_2}h_{\ell_3}  \mp \frac{1}{8}e^{\Phi}\tilde{\nabla}_{\ell_1}\Phi X_{\ell_2 \ell_3}
- \frac{1}{24}e^{\Phi}\tilde{\nabla}^{i}\Phi \tilde{G}_{i \ell_1 \ell_2 \ell_3}
\nonumber \\
&\mp& \frac{1}{32}e^{\Phi}\tilde{F}_{\ell_1 \ell_2}L_{\ell_3}
\mp \frac{1}{96}e^{\Phi}S\tilde{H}_{\ell_1 \ell_2 \ell_3} - \frac{1}{32}e^{\Phi}\tilde{F}^{i}{}_{\ell_1}\tilde{H}_{i \ell_2 \ell_3}\big)\Gamma^{\ell_1 \ell_2 \ell_3}
\nonumber \\
&+& \Gamma_{11}\bigg(\big(\mp \frac{1}{4}\tilde{\nabla}_{\ell}(e^{\Phi}S) - \frac{1}{4}\tilde{\nabla}^{i}(e^{\Phi}\tilde{F}_{i \ell}) +\frac{1}{16}e^{\Phi}S h_{\ell} \pm \frac{1}{16}e^{\Phi}h^{i}\tilde{F}_{i \ell} \pm \frac{1}{4}e^{\Phi}\tilde{\nabla}_{\ell}\Phi S
\nonumber \\
&+& \frac{1}{4}e^{\Phi}\tilde{\nabla}^{i}\Phi\tilde{F}_{i \ell} + \frac{1}{16}e^{\Phi}L^i X_{i \ell}
\mp \frac{1}{32}e^{\Phi}\tilde{H}^{i j}{}_{\ell}X_{i j}
- \frac{1}{96}e^{\Phi}\tilde{G}^{i j k}{}_{\ell}\tilde{H}_{i j k}\big)\Gamma^{\ell}
\nonumber \\
&+& \big(\mp \frac{1}{4}\tilde{\nabla}_{\ell_1}(L_{\ell_2}) - \frac{1}{8}\tilde{\nabla}^{i}(\tilde{H}_{i \ell_1 \ell_2}) + \frac{1}{4}\tilde{\nabla}^{i}\Phi \tilde{H}_{i \ell_1 \ell_2} \pm  \frac{1}{16}h^{i}\tilde{H}_{i \ell_1 \ell_2}\big)\Gamma^{\ell_1 \ell_2}
\nonumber \\
&+& \big(\pm \frac{1}{384}e^{\Phi}\tilde{G}_{\ell_1 \ell_2 \ell_3 \ell_4}L_{\ell_5} \pm  \frac{1}{192}e^{\Phi}\tilde{H}_{\ell_1 \ell_2 \ell_3}X_{\ell_4 \ell_5}
+ \frac{1}{192}e^{\Phi}\tilde{G}^{i}{}_{\ell_1 \ell_2 \ell_3}\tilde{H}_{i \ell_4 \ell_5}\big)\Gamma^{\ell_1 \ell_2 \ell_3 \ell_4 \ell_5}\bigg)
\bigg] \eta_\pm
\nonumber \\
&+& {1 \over 2} \big(1 \mp 1\big) \bigg(h^i {\tilde{\nabla}}_i \Phi
-{1 \over 2} {\tilde{\nabla}}^i h_i \bigg) \eta_\pm \ .
\eea

Note that with the exception of the final line of the RHS of ({\ref{quad}}), all terms on the RHS of the above expression
give no contribution to the second line of (\ref{laplacian}),
using (\ref{hermiden1}) and (\ref{hermiden2}), since all these terms in (\ref{quad}) are anti-Hermitian and thus the bilinears vanish.
Furthermore, the contribution to the Laplacian of $\parallel \eta_+ \parallel^2$ from the final line of ({\ref{quad}}) also vanishes;
however the final line of ({\ref{quad}}) {\it does} give a contribution
to the second line of ({\ref{laplacian}}) in the case of the
Laplacian of $\parallel \eta_- \parallel^2$.
We  proceed
to consider the Laplacians
of $\parallel \eta_\pm \parallel^2$ separately, as the analysis
of the conditions imposed by the global properties of ${\cal{S}}$
differs slightly in the two cases.

For the Laplacian
of $\parallel \eta_+ \parallel^2$, we obtain from ({\ref{laplacian}}):
\bea
\label{l1}
{\tilde{\nabla}}^{i}{\tilde{\nabla}}_{i}\parallel\eta_+\parallel^2 - (2\tilde{\nabla}^i \Phi +  h^i) {\tilde{\nabla}}_{i}\parallel\eta_+\parallel^2 = 2\parallel{\hat\nabla^{(+)}}\eta_{+}\parallel^2 - (4\kappa + 16 \kappa^2)\parallel\mathcal{A}^{(+)}\eta_+\parallel^2 \ .
\nonumber \\
\eea
This proves (\ref{maxprin}).

The Laplacian of $\parallel \eta_- \parallel^2$
is calculated from ({\ref{laplacian}}), on taking account of the contribution to the second line of
({\ref{laplacian}}) from the final line of ({\ref{quad}}). One
obtains
\bea
\label{l2}
{\tilde{\nabla}}^{i} \big( e^{-2 \Phi} V_i \big)
= -2 e^{-2 \Phi} \parallel{\hat\nabla^{(-)}}\eta_{-}\parallel^2 +   e^{-2 \Phi} (4 \kappa +16 \kappa^2) \parallel\mathcal{A}^{(-)}\eta_-\parallel^2~,
\nonumber \\
\eea
where
\bea
V=-d \parallel \eta_- \parallel^2 - \parallel \eta_- \parallel^2 h \ .
\eea
This proves (\ref{l2b}) and completes the proof. 

It should be noted that in the $\eta_-$ case, one does not have to set $q=-1$. In fact, a formula similar to (\ref{l1}) can be established 
for arbitrary $q$. However some terms get modified and the end result does not have the simplicity of (\ref{l1}).  For example, 
 the numerical coefficient in front of the $\parallel\mathcal{A}^{(-)}\eta_-\parallel^2$ is modified to
$-2-4\kappa q+ 16 \kappa^2+2q^2$ and of course reduces to that of (\ref{l1}) upon setting  $q=-1$.

\end{document}